# Aggregation of hydrophobic-amphiphilic block copolymers

Sophia A Pavlenko[1], Elena N Govorun[2*]

[1] *Faculty of Physics, Lomonosov Moscow State University, 119991 Moscow, Russia*

[2] *UMR CNRS 7083 Gulliver, ESPCI Paris, PSL Research University, 75005 Paris, France*
elena.govorun@espci.fr, elenagovorun653@gmail.com

We analyze the aggregation of locally amphiphilic copolymers with blocky architecture and uniformly distributed amphiphilic moieties in terms of a mean-field theory. Locally amphiphilic structure is characteristic of many thermoresponsive polymers, both linear and grafted, which endows them with local surface activity. Self-assembly of such copolymers exhibits a rich diversity of morphologies, which are analyzed in the present work in the limit of high surface activity of amphiphilic dimers with solvophilic/polar pendants. Depending on the composition and architecture of the copolymer and sizes and interaction parameters of the solvophilic/polar pendants, we build morphological diagrams of copolymer solutions. Copolymers with small volume pendants form precipitates or large aggregates with internal voids containing these pendants. For moderate volume pendants, a lamellar structure (or large vesicles) is observed at smaller fractions of amphiphilic monomer units in the chain and micelles are formed at larger fractions of these units. The sizes and shapes of micelles depend on the monomer distribution along the chain, and the blocky architecture favors the existence of single spherical micelles or spherical particles that constitute compound micelles. For sufficiently large polar pendants, copolymers of both types mainly form granular branched compound micelles. The distance at which adjacent hydrophobic and polar groups are located in a chain determines the size of domains and particles formed by locally amphiphilic blocks or chains. Our predictions explain the existence of dense multicore complex micelles consisting of beads of several tens of nanometers size and other structures such as disc-shaped micelles.

# 1. Introduction

Morphological variability of locally amphiphilic polymers, which are mostly thermoresponsive, is very rich. They possess local surface activity favorable for the formation of aggregates with high surface-to-volume ratio. In particular, many synthetic amphiphilic homopolymers, amphiphilic graft copolymers and natural polymers form various types of particles in solutions, such as disc-shaped micelles, vesicles, mesoglobules and compound micelles.[1-14] Understanding the principles of self-assembly is necessary in order to control the structure of the aggregates and their ability to carry molecular cargo such as drugs or catalysts, or to reproduce the elements of living cells.

Both amphiphilic homopolymers, where adjacent hydrophobic (or solvophobic) and polar (or solvophilic) moieties are separated by a distance of the order of monomer unit size, and graft copolymers, where this distance is larger, can exhibit the local surface activity. To this end, the mutual orientation of the hydrophobic and polar groups must be variable due to chain flexibility or the rotational freedom of the pendants. For example, such homopolymers as poly(*N*-isopropylacrylamide) (PNIPAM) and amphiphilic graft copolymers promote micellization into various morphologies,[1-14] the characteristics of which depend on the types of monomer units and sizes of their constituent parts. Polymers consisting of locally amphiphilic parts linked to hydrophobic or hydrophilic blocks, offer even more opportunities for self-assembly,[15-25] with hydrophobic blocks being considered as additional agents capable of binding to hydrophobic substances. The self-assembly is also affected by the external conditions such as temperature and concentration of solutes, in particular, chaotropic salts or surfactants.[8,13,26-28] The salt ions can be bound to the polymer groups by non-covalent interactions, thus creating additional pendants in the polymer chains and promoting the local surface activity of macromolecules.

When the detailed structure of micelles of locally amphiphilic macromolecules was characterized experimentally, in particular with the use of electron microscopy, large branched aggregates were often observed; with the size of the constituting particles not exceeding several tens of nanometers. They were described as "a network of spherical particles" or "multicore micelles" for the amphiphilic graft copolymer self-assembling during the polymerization reaction[4] and as "aggregates with a characteristic hydrophobic multicore structure" for a copolymer with two types of grafted chains.[7] Amphiphilic random copolymers formed "aggregates of the core-shell spherical micelle subunits".[9] Different block copolymers self-assembled into "compound micelles",[19-21] "micellar aggregates",[22] and even crosslinked "multicompartement worms".[24] It is not easy to reveal the role of macromolecular architecture, as well as of the lengths and volumes of the constituents, in the solution morphology in experiments and in computer simulations. However, a wide range of parameters can be easily addressed using theoretical models, thus facilitating control of the micelle structure.

We study aggregation of block copolymers with hydrophobic/solvophobic and amphiphilic blocks depending on the chain length and composition, the length and volume of pendants, and the Flory-Huggins interaction parameters. Locally amphiphilic segments of the macromolecules are described as dimers with hydrophobic/solvophobic groups in the main chain and polar/solvophilic groups as pendants. Favorable shapes and sizes of micelles in solution and also voids in precipitates (or structured agglomerates) are determined by analyzing and comparing the free energies of different morphologies. We consider architectures of block copolymers, for which the hydrophobic blocks usually form the cores of micelles, and copolymers (homopolymers) with uniformly distributed amphiphilic monomer units, which are assumed to form homogeneous micelles.

First, micelle morphologies of the polymers (amphiphilic homopolymers) with dimer, or "dumbbell", monomer units with hydrophobic main chain and polar pendants were studied for single macromolecules, both in computer simulations[29] and theoretically,[30] and these considerations were later generalized to a wider range of parameters and a more detailed model for free energy calculations.[31] A rich set of shapes, including spheres, necklace-like conformation, vesicles, cylinders, and toroidal micelles, was described depending on the chain length, solvent quality, and second virial coefficient of interactions between hydrophobic and polar groups.[31] In the theoretical investigation of a collapsed single copolymer molecule with a hydrophobic block and an amphiphilic block consisting of dimers,[25] a spherical core-shell particle, disc, and tadpole conformation of a core-shell particle connected to a necklace of beads were predicted.

In molecular dynamic simulations of solutions of amphiphilic homopolymers with hydrophobic backbones and polar pendants, necklace-like and cylindrical conformations were observed,[32] whereas the macromolecules with polar backbone and hydrophobic pendants formed spherical, cylindrical or toroidal particles and vesicles, sometimes of open-mouth or perforated types.[33]

The solution morphology of polymers with uniformly distributed amphiphilic monomer units represented as dimers with polar pendants was analyzed theoretically by Maresov and Semenov in relation to the elastic moduli of the surface of a dense condensed polymer.[34] Such structures as spheres, gyroid, cylinders (and inverse micelles), and macrophases were predicted for positive values of the mean bending modulus and layers/vesicles for its negative values. Earlier, the micellization of polymers with polar main chains and hydrophobic pendants (polysoaps) was considered by Borisov and Halperin[35] and comb-like copolymers with hydrophobic main chains and polar side chains[36] by Borisov and Zhulina using scaling theory. It was found that necklaces of intrachain micelles were mostly formed, the sizes of which were determined mainly by the conformational free energies of swelling of the chains or spacers in polar coronas and hydrophobic cores. Cylindrical and disc-like micelles were predicted for sparse grafting of hydrophobic side chains.[36]

In the present work, we generalize our previous theoretical considerations of micelles of single diblock copolymer molecules with hydrophobic and amphiphilic blocks[25] and single macromolecules with uniformly distributed polar pendants[31] to micelles of arbitrary aggregation numbers and to macromolecules with different lengths pendant. It is assumed that the main chains are highly hydrophobic and the amphiphilic monomer units are characterized by high surface activity so that dense micelles are formed with the polar groups exposed to solvent. These conditions lead to a strong effect of the distance between adjacent hydrophobic and polar moieties in the chain (a pendant length) on the size of the amphiphilic block domains.

In our consideration, the pendant length is fixed, so that the model is relevant for macromolecules with moderate lengths of side chains or pendants. In this case, shape and size of micelles are determined mainly by the interaction energies rather than chain stretching. Besides, we consider disc-like and cylindrical micelles of finite size, which are transformable into spheres under the continuous change of their geometrical parameters.

The chain architectures, considered structures and the free energy of spherical micelles are described in *Model* section, the calculations for the other micelle types and precipitates are detailed in *Supporting information* (after the main text). The morphological diagrams are presented in *Results and discussion* section, where the morphologies of block copolymers are analyzed in comparison with those of polymers with uniformly distributed amphiphilic monomer units. The results are summarized in *Conclusions*.

## 2. Model

In solution of block copolymers with hydrophobic (solvophobic) and thermoresponsive blocks, micelles of different shapes and sizes can be formed depending on the molecular parameters and temperature. We compare the micelle morphologies of block copolymers with those of regular copolymers (Fig. 1) of the same mean composition and also with the shapes of single collapsed macromolecules. For a diblock copolymer, an amphiphilic block consists of $N_A$ dimer monomer units and a hydrophobic block consists of $N_H$ hydrophobic monomer units ($H$ beads). The amphiphilic blocks contain hydrophobic groups in the main chain and polar groups as pendants. By altering the size of the pendant, we can describe both polymers with small pendants and bottle-brush, or grafted, polymers using the dimer model. The main chain length is $N = N_H + N_A$. In a regular copolymer, the same amphiphilic monomer units are uniformly distributed along the chain.

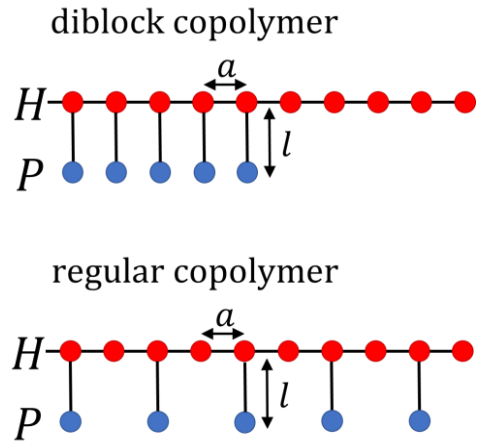

Figure 1. Macromolecule structures of a diblock copolymer with amphiphilic and hydrophobic blocks and a copolymer with regularly and uniformly distributed amphiphilic monomer units. Hydrophobic (*H*) and polar (*P*) groups are shown in red and blue, respectively.

The volumes of hydrophobic and polar groups are denoted by $\upsilon_H$ and $\upsilon_P$, respectively, the volume of solvent molecule by $\upsilon_S$. The monomer unit size along the backbone chain is equal to $a$, the length of bond between *H* and *P* groups is $l$. The polymer chains are assumed to be flexible and the side bonds with *P* groups are able to freely rotate around the main chain. The hydrophobicity of the *H* groups is determined by the value of the Flory-Huggins parameter $\chi_{HS}$ describing the *H* group interactions with solvent. The Flory-Huggins parameter $\chi_{HP}$ describes the *H* and *P* group interactions that control the surface activity of amphiphilic monomer units. The polar groups tend to be surrounded by solvent, and the zero value of the Flory-Huggins parameter $\chi_{PS} = 0$ is taken.

The repulsion between hydrophobic groups and solvent leads to the formation of micelles, whereas the repulsion between hydrophobic and polar groups can lead to the segregation of hydrophobic and amphiphilic blocks. Denote the micelle volume by *V* and the volume of the core of hydrophobic blocks by $V_0$. If amphiphilic blocks are collapsed, the *H-P* bonds at the surface are oriented so that *P* groups are exposed to the solvent.

Previously, shapes and sizes of collapse macromolecules were studied by us for single macromolecules of diblock copolymers and macromolecules with regularly distributed amphiphilic monomer units in solution[25,31] that can form particles of tens of nanometers maximum size. In the present work, we generalize this approach for micelles, such as mesoglobules or complex micelles, of arbitrary sizes and for precipitates, and additionally study the effect of the side bond/chain length on solution morphology. More structure types of micelles are analysed (Fig. 2): complex, or compound, micelles consisting of interconnected particles (CM), isolated spherical particles (IS), disc-like micelles (D), cylindrical micelles (C), and vesicles (V). It is assumed that a smoothed cylinder and disc can continuously transform into a sphere under decreasing their height and its radius, respectively. Micelles with cores of segregated hydrophobic blocks and homogeneous micelles formed by macromolecules with short hydrophobic blocks and with regularly distributed amphiphilic monomer units are considered.

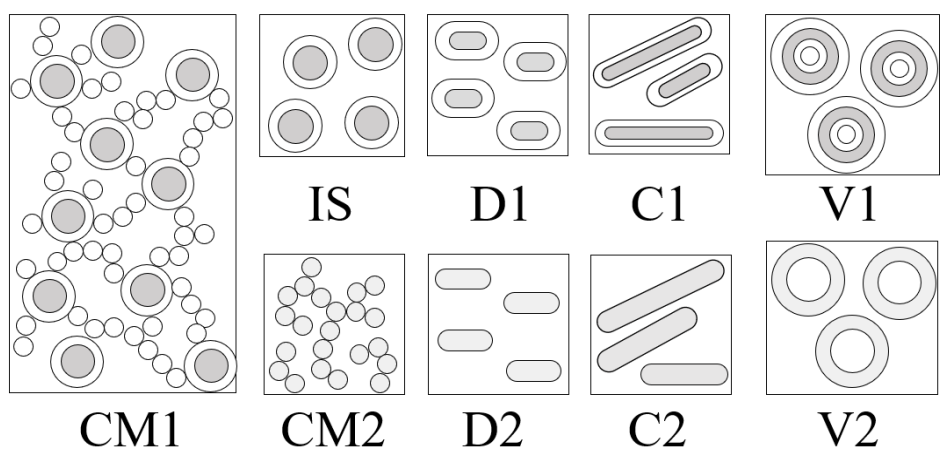

Figure 2. Solution morphologies: CM - complex micelles, IS - isolated spheres, D - disc-shaped micelles, C - cylindrical micelles, V - vesicles. The micelles with segregated cores of hydrophobic blocks (CM1, IS, D1, C1, V1) and homogeneous micelles (CM2, D2, C2, V2) are considered.

These micelles can be described similarly to the globules of single macromolecules[25,31] with the additional consideration of the aggregation number. Besides, both block copolymers and regular copolymers are expected to precipitate (or form large aggregates), if the fraction of amphiphilic monomer units is quite small. In the precipitate, polar side groups tend to segregate from hydrophobic groups, thus leading to the formation of cavities containing only polar groups in the solvent. We analyze the homogeneous precipitate (Phom), precipitate with a lamellar structure (Pl), and precipitates with cylindrical and spherical cavities ("inverse micelles") as possible morphologies (Fig. 3).

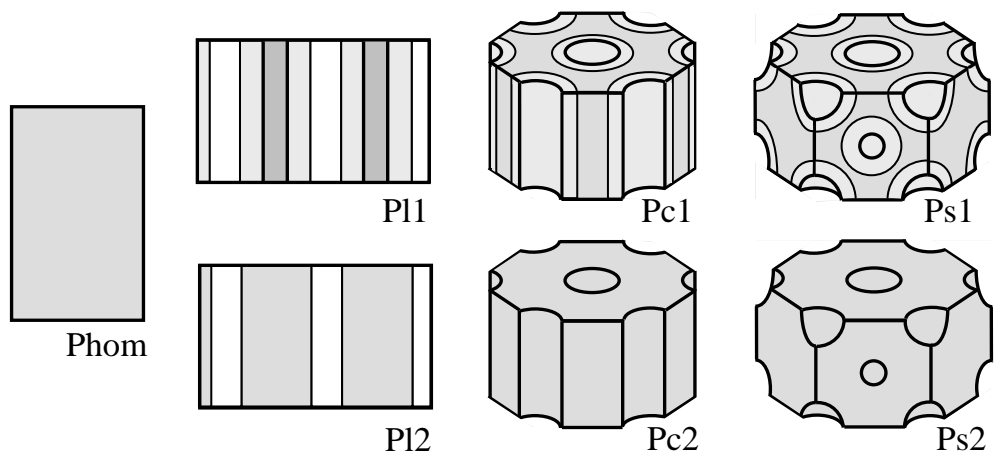

Figure 3. Precipitate morphologies: Phom – homogeneous precipitate, Pl – lamellar precipitate, Pc – precipitate with hexagonally arranged cylindrical cavities, Ps – precipitate with spherical cavities at the vertices of a hexagonal close-packed lattice. Hydrophobic and amphiphilic blocks are in different domains for the morphologies Pl1, Pc1, and Ps1. The condensed polymer is homogeneous for the morphologies Phom, Pl2, Pc2, and Ps2.

For each morphology, the most favourable particle or cavity sizes and the corresponding free energy value are calculated. The minimum free energy per monomer unit among all structure types (Figs. 2 and 3) determines the observed morphology presented in the diagrams (*Results and discussion* section).

The domain of hydrophobic blocks and domain of amphiphilic blocks are assumed to be homogeneous and the interfaces between the core and the shell and between the shell and the outer solvent are assumed to be thin, so that the interface thickness is of the order of the monomer unit size $a$ and the *H-P* bond length $l \gg a$. It is also assumed that in the considered case of high surface activity of amphiphilic monomer units, all hydrophobic groups of these units in the surface layer of maximum thickness $l$ have their polar groups outside in the outer adjacent layer of thickness $l$.

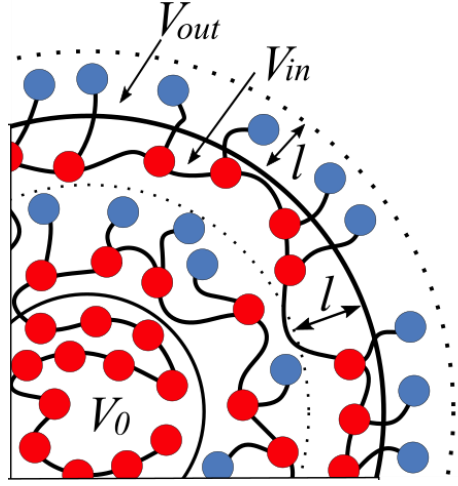

Figure 4. Schematic of a micelle with the core of hydrophobic blocks and the shell of amphiphilic blocks. The micelle surface (thick black curve) comprises all hydrophobic beads. The hydrophobic groups of amphiphilic monomer units in the surface layer of thickness $l$ have their polar groups outside in the outer adjacent layer of thickness $l$.

In this section, we describe in detail the spherical core-shell micelles (Fig. 4) of radius $R$ with the core of radius $R_0$, which consist of $m$ macromolecules each. The volume of the surface layer of thickness $l$ is denoted by $V_{in}$ and the volume of the outer adjacent layer of thickness $l$ by $V_{out}$. Let $\varphi_{H0}$ and $\varphi_H$ be the volume fractions of hydrophobic groups, $\varphi_{S0}$ and $\varphi_S$ be the solvent volume fractions in the homogeneous micelle core and shell, respectively. Let also $\varphi_P$ and $\psi_P$ be the volume fractions of polar groups in the shell and in the outer adjacent layer. The incompressibility conditions are: $\varphi_{S0} + \varphi_{H0} = 1$ and $\varphi_S + \varphi_H + \varphi_P = 1$. The core-shell structures are analyzed only for core radius and shell thickness greater than the mean diameter of the hydrophobic group $d_H = (6\upsilon_H/\pi)^{1/3}$.

The volume fractions of hydrophobic groups for a core-shell micelle can be written as

$$\varphi_{H0} = \frac{mN_H \upsilon_H}{V_0}, \quad \varphi_H = \frac{mN_A \upsilon_H}{V - V_0}. \tag{1}$$

The volume fractions of polar groups are:

$$\varphi_P = \frac{N_{Pin}\upsilon_P}{V - V_0}, \quad \psi_P = \frac{N_{Pout}\upsilon_P}{V_{out}}, \tag{2}$$

where $N_{Pin}$ and $N_{Pout}$ are the numbers of polar groups in the shell and in the outer adjacent layer, $N_{Pin} + N_{Pout} = mN_A$. If the thickness of the shell of amphiphilic blocks $\Delta R = R - R_0 < l$, then $N_{Pout} = mN_A$ and if $\Delta R \geq l$, then $N_{Pout} = \varphi_H V_{in}/\upsilon_H$, where the volume of the inner adjacent layer is $V_{in} = 4\pi l(R^2 - Rl + l^2/3)$.

The core-shell micelle free energy $F_{c-s}$ consists of the free energies of volume interactions in the core and in the shell, $F_{bulk0}$ and $F_{bulk}$, the free energy $F_{inter}$ of a core-shell interface, the surface free energy $F_{surf}$, and the contribution of volume interactions in the outer adjacent layer, $F_{layer}$:

$$F_{c-s} = F_{bulk0} + F_{bulk} + F_{inter} + F_{surf} + F_{layer} \tag{3}$$

These free energy terms are the same as for a globule formed by a single macromolecule,[21] however, in the present approach an arbitrary aggregation number and micelle sizes can be considered. All free energy contributions below are written out per monomer unit (or per hydrophobic group).

The core free energy consists of the pair interaction energy of the hydrophobic groups with the solvent and the entropic contribution of polymer-solvent mixing:

$$\frac{F_{bulk0}}{mNk_BT} = (1-f)\left(\chi_{HS}\varphi_{S0} + \frac{\upsilon_H}{\upsilon_S}\frac{\varphi_{S0}}{\varphi_{H0}}\ln\varphi_{S0}\right), \tag{4}$$

where $f = N_A/N$ and $(1-f)$ are the length fractions of amphiphilic and hydrophobic blocks in a macromolecule, $k_B$ is the Boltzmann constant, $T$ is the thermodynamic temperature. It is used that $\varphi_{S0}V_0/\upsilon_S$ is the number of solvent molecules in the core.

The free energy of the volume interactions in the shell is

$$\frac{F_{bulk}}{mNk_BT} = f\left(\chi_{HP}\varphi_P + \chi_{HS}\varphi_S + \frac{\upsilon_H}{\upsilon_S}\frac{\varphi_S}{\varphi_H}\ln\varphi_S\right). \tag{5}$$

The free energy of the core-shell interface contains the contribution of the elastic free energy of polymer chain deformation at the interface and the pair interaction term:

$$\frac{F_{inter}}{mNk_BT} = \frac{1}{mN}\left(\frac{S_0}{\upsilon_S^{2/3}}\gamma_0(\varphi_{H0} - \varphi_H) + \frac{E_{inter}}{k_BT}\right). \tag{6}$$

Here $S_0 = 4\pi R_0^2$ is the interface area between the core and the shell, $\gamma_0$ is the reduced value of the interfacial tension (of the order of unity) describing the elastic free energy per excess monomer unit in the core interfacial layer relative to the shell. The number of such monomer units is estimated as $N_{inter} = (\varphi_{H0} - \varphi_H) S_0/v_S^{2/3}$, $\varphi_{H0} > \varphi_H$.

The pair interaction contribution $E_{inter}$ and the surface free energy $F_{surf}$ are taken in the lattice model:[25,31]

$$\frac{E_{inter}}{mNk_BT} = \frac{S_0}{2mNv_S^{2/3}}(\varphi_{H0} - \varphi_H)\left((\chi_{HP} - \chi_{HS})\varphi_P + \chi_{HS}(\varphi_{H0} - \varphi_H)\right), \qquad (7)$$

$$\frac{F_{surf}}{mNk_BT} = \frac{S}{mNv_S^{2/3}}\varphi_H\left(\gamma_0 + \frac{1}{2}\left((\chi_{HP} - \chi_{HS})(\psi_P - \varphi_P) + \chi_{HS}\varphi_H\right)\right), \qquad (8)$$

where $S = 4\pi R^2$ is the micelle surface area.

In the free energy of the outer layer, only the translational entropy of the solvent molecules is taken into account:

$$\frac{F_{layer}}{mNk_BT} = \frac{V_{out}}{mNv_S}(1 - \psi_P)\ln(1 - \psi_P), \qquad (9)$$

where $V_{out} = 4\pi l(R^2 + Rl + l^2/3)$ is the volume of the outer adjacent layer of thickness $l$. This term describes, in fact, the excluded volume interactions of polar groups outside the micelle.

With the normalization conditions and the geometrical relations for spherical micelles (IS), the free energy $F_{c-s}$ (3) can be considered as a function of three independent variables, and it is convenient to use the volume fractions $\varphi_{H0}$ and $\varphi_H$ and the aggregation number $m$ as such variables. By minimizing the free energy per monomer unit, $F_{c-s}/(mN)$, with respect to these three variables, we determine the equilibrium values of the micelle parameters and the free energy of the morphology of spherical particles.

For other morphologies, more geometric parameters need to be used. Expressions describing the core-shell cylindrical micelles, disc-shaped micelles, vesicles, necklaces of spherical particles (complex or compound micelles), and various types of precipitates (Figs. 2 and 3) can be found in *Supporting information*. To describe the precipitates with cavities, we consider the ordered arrangement of cylindrical cavities in a hexagonal lattice and a hexagonal close-packed lattice for spherical ones. Such types of arrangement make possible a high density of cavities that is expected to be favorable to provide a large surface area of the condensed polymer. Morphologies of micelles and precipitates are determined for both block copolymers and regular copolymers to reveal the role of polymer architecture.

The observed morphology corresponds to the free energy minimum among the considered structure types under the given model parameters. Typical sizes of the micelles and distances between cavities do not exceed the unperturbed sizes of the blocks, therefore the blocks are not stretched and their conformational energies don't play a role in the bulk free energies of homogeneous domains, they are taken into account only in surface free energies. In terms of this model, the free energy of copolymers with blocky architecture is independent of the number of blocks, if the blocks are long enough to form the core and shell domains, and the morphological diagrams can describe copolymers with different numbers of blocks.

Morphological diagrams of solutions of regular copolymers with a small fraction of amphiphilic monomer units were earlier analyzed by Maresov and Semenov,[34] and we compare our predictions for such copolymers with those of Ref. 34 in *Results and discussion* section. However, it is worth noting that the model assumptions are significantly different. We consider only the limit of very high surface activity of amphiphilic monomer units, when all capable polar groups are pushed out of the condensed polymer. At the same time, we take into account that the bulk properties and the bulk free energy of micelles (mesoglobules), or precipitates, depend on the amount of polar groups outside, whereas the bulk free energy was assumed to be constant in Ref. 34. We also explicitly consider the effect of volume interactions of polar groups in the outer adjacent layer of the micelles.

## 3. Results and discussion

Solution morphology is highly dependent on macromolecular architecture, in particular, on the amphiphilic block fraction $f = N_A/N$ and the side group volume $\upsilon_P$. First, we calculate the typical structures successively for small, moderate, and large values of the volume $\upsilon_P$ depending on the amphiphilic block fraction, chain length, and other parameters.

In Fig. 5, the diagrams for the small volume of side groups ($\upsilon_P = 0.6\upsilon_S$) are presented in the coordinates (a) $f$, $N$ and (b) $l$, $f$. The macromolecules form a precipitated phase or macroscopic aggregates with segregated hydrophobic and amphiphilic blocks. The morphological type of the internal structure depends mostly on the fraction $f$ of amphiphilic blocks rather than the chain length or the length of side bonds. With increasing $f$, the separation of polar groups in spherical (Ps1), then cylindrical (Pc1), and finally lamellar cavities (Pl1 and Pl2) are observed. The thickness of the amphiphilic layers is close to $l$ for all cavity shapes to ensure that all polar groups can be in the cavities. Within each region at the diagram, the sizes of the cavities and amphiphilic layers are almost constant, whereas the distance between cavities decreases with increasing $f$. When the layers of amphiphilic blocks become almost overlapping, the spherical cavities transform to cylindrical ones, or the cylindrical cavities transform to lamellas. The free energies of lamellar precipitate and large vesicles are virtually equal to each other and these morphologies are combined into a common Pl/V region on the diagram. For very large amphiphilic blocks (the region separated by a dotted line), the hydrophobic blocks are intermixed with the amphiphilic ones and do not form separate domains (Pl2 or V2).

For longer side chains (larger $l$), the transitions from the precipitate with spherical cavities (Ps1) to the precipitate with cylindrical ones (Pc1), and from Pc1 to the lamellar precipitate (Pl1) occur at slightly smaller $f$ (Fig. 6b). Both the thickness of the amphiphilic layers and the size of the cavities are approximately determined by the length $l$. For small volumes of $P$ groups, the value of the volume fraction of polar groups in the outer layer $\psi_P$ is quite small and the change in the lateral pressure of the $P$ groups with $l$ has virtually no effect on the transitions. At the same time, for a fixed fraction $f$, the cavity concentration scales with this length as $1/l^\alpha$, where $\alpha = 3$ for spherical, 2 for cylindrical, and 1 for planar cavities. Then, the surface area of cavities of any shape is inversely proportional to $l$, and the increase in surface free energy upon the transitions Ps1→Pc1 or Pc1→Pl1 is smaller for larger $l$. This facilitates the transitions at larger $l$, resulting in slightly smaller transition values of the fraction $f$.

The decrease in $\chi_{HS}$ (improvement of the solvent quality) leads to swelling of hydrophobic domains and to the surface tension drop. Correspondingly, the spatial scales of the structures increase and the layers of amphiphilic blocks begin to overlap at smaller fractions $f$ of amphiphilic blocks with increasing $f$. As a consequence, the morphological transitions occur at smaller values of $f$ (compare the red and black curves in Fig. 5).

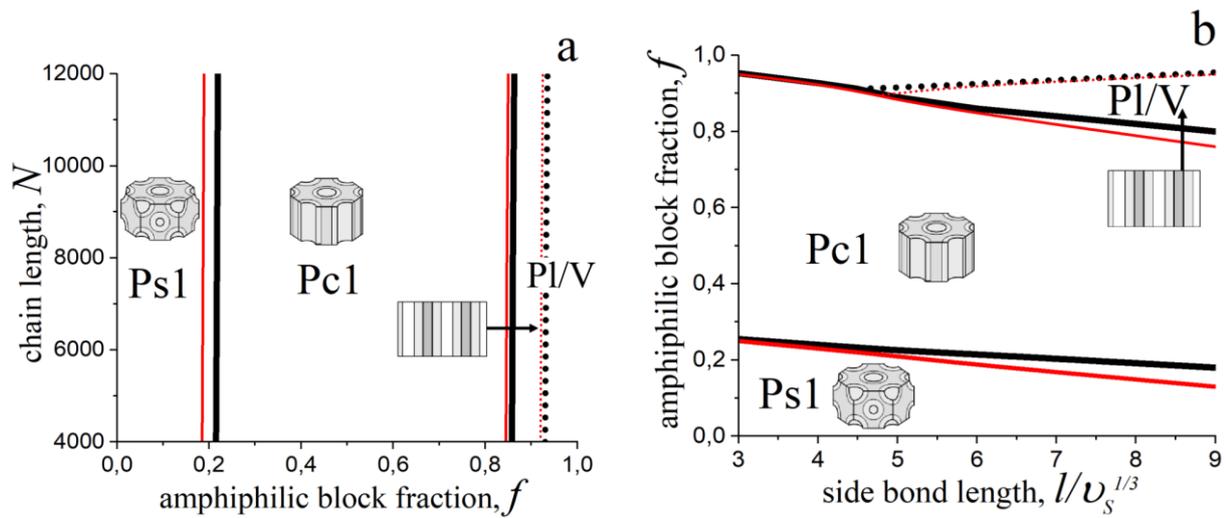

Figure 5. Morphological diagrams of block copolymers with amphiphilic blocks with a small volume of the polar groups, $\upsilon_P/\upsilon_S = 0.6$, in the coordinates (a) $f$ (the fraction of the amphiphilic block) – $N$ (the chain length) for $l = 6\upsilon_S^{1/3}$ and (b) $l$ (the side bond length) – $f$ for $N = 8000$. The Flory-Huggins parameters: $\chi_{HP} = 4.4$, $\chi_{HS} = 4.2$ (thick black curves) and 3.8 (thin red curves). Other parameters: $\upsilon_H/\upsilon_S = a/\upsilon_S^{1/3} = 1.5$, $\gamma_0 = 0.8$.

The solution morphology of copolymers with regularly distributed amphiphilic monomer units was additionally analyzed. With the same parameters as in Fig. 5a, calculations predict condensation of the polymer into a precipitate (or large aggregates): at the fraction of amphiphilic monomer units $f < 0.2$, the precipitate is homogeneous, and at $f > 0.2$ it has a layered structure (or large vesicles can be formed). Spherical and cylindrical cavities are unfavorable, since for them the concentration of cavities must be higher to ensure that all polar groups can be in the cavities. This would carry a penalty in surface energy and prevent the formation of finite cavities.

For block copolymers with larger volume of polar groups ($v_P/v_S = 1.5$), a lamellar precipitate (Pl1 or V1) is observed at fairly small fractions $f$ of amphiphilic blocks, and the polymer forms micelles at larger fractions $f$ (Fig. 6). For more voluminous $P$ groups, their steric repulsion in the outer layers plays a more important role. Cavities with concave surfaces in the condensed polymer are not formed and, with increasing fraction $f$, the precipitate loses its stability and large spherical micelles (IS) first appear. Their volume $V_{\text{out}}$ of the outer adjacent layer is significantly greater than that of the lamellar precipitate that reduces the lateral pressure of $P$ groups in this layer and makes the spherical morphology favourable. As $f$ increases, the number of $P$ groups and the outer volume $V_{\text{out}}$ increase. The thickness of the layer of amphiphilic blocks $\Delta R$ is close to $l$, whereas the volume fraction of this layer in the whole micelle is approximately equal to $f \approx S\Delta R/V$ for large micelles with $\varphi_H \approx \varphi_{H0}$. This leads to a decrease in radius of spheres and thickness of discs with increasing $f$ (Fig. 7). For spheres $S/V = 3/R$ and their radius is approximately inversely proportional to $f$, $R \sim 3/f$.

In parallel, the lateral pressure in the outer adjacent layer leads to swelling of micelles and, thus, to an increase in the free energy of volume interactions. This effect is most pronounced for small spherical micelles with a high $S/V$ ratio. As a result, macromolecules with long amphiphilic blocks form flattened micelles of a smoothed disc-like shape (D1 in Figs. 6, 7). This change in the shape of small micelles allows maintaining the same free energy of the outer layer as for the spherical micelles, but reducing the bulk interaction energy. If the hydrophobic blocks are very short, they are homogeneously distributed in the discs without forming cores (D2).

In the morphological diagram in the coordinates $l - f$ (Fig. 6b), all morphological transitions require a higher fraction $f$ of amphiphilic blocks with increasing length $l$. At a fixed fraction $f$, the size of the micelles and the thickness of the precipitated layers increase with $l$ simultaneously with the thickness of the amphiphilic block layers. All $P$ groups are outside and its volume fraction $\psi_P$ is determined by the total volume of the adjacent layer. The increase in this volume at the transition precipitate-spherical micelles is greater for smaller micelles due to their greater curvature, resulting in a transition at smaller $f$. Smaller micelles swell more, which promotes the transition to flattened micelles (D1) also at smaller $f$.

The decrease in $\chi_{HP}$ corresponds to a decrease in surface activity of amphiphilic monomer units. For a purely hydrophobic particle, only the surface free energy decreases (Eq. (8)) and it becomes less sensitive to the value of the volume fraction of polar groups $\psi_P$ in the outer adjacent layer. The transitions from lamellar structure (or vesicles) to core-shell spheres and further to disc-like micelles occur at larger fractions $f$ (compare the blue and black solid lines in Fig. 6). However, with increasing fraction $f$ the discs become homogeneous (D2) at smaller $f$, since the contacts between $H$ and $P$ groups in the homogeneous micelles become less unfavourable.

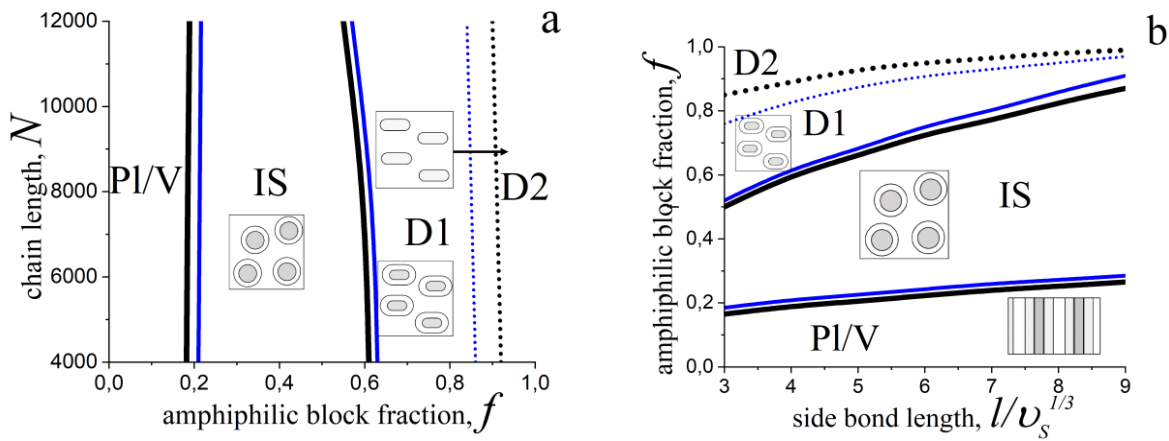

Figure 6. Morphological diagrams of block copolymers with a moderate volume of the polar groups, $\upsilon_P / \upsilon_S = 1.5$, in the coordinates (a) $f - N$ for $l = 3.75 \upsilon_S^{1/3}$ and (b) $l - f$ for $N = 8000$. The Flory-Huggins parameters: $\chi_{HS} = 4.2$, $\chi_{HP} = 4.4$ (thick black curves) and 3.4 (thin blue curves). Other parameters: $\upsilon_H / \upsilon_S = a / \upsilon_S^{1/3} = 1.5$, $\gamma_0 = 0.8$.

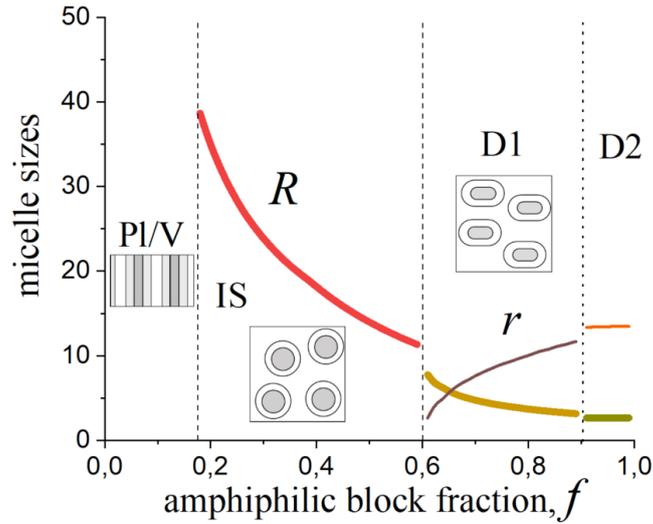

Figure 7. Sizes of block copolymer micelles in units of $a$ ($R$ is the radius of spheres and the half-thickness of discs, $r$ is the radius of discs) depending on the fraction $f$ of amphiphilic blocks for a moderate volume of polar groups, $\upsilon_P/\upsilon_S = 1.5$. Other parameters: $N = 8000$, $l = 3.75\upsilon_S^{1/3}$ $\upsilon_H/\upsilon_S = a/\upsilon_S^{1/3} = 1.5$, $\gamma_0 = 0.8$, $\chi_{HP} = 4.4$, $\chi_{HS} = 4.2$.

For copolymers with a uniform distribution of amphiphilic monomer units in the chains with the same parameters as in Fig. 6a, at $f < 0.6$ a homogeneous (Phom) or layered (Pl2 or V2) precipitate is formed and at higher $f$ disc-like micelles (D2) are formed (Fig. 8a). The layer thickness in the precipitate is close to $2l$ allowing all P groups to be outside. As $f$ increases, the volume fraction of P groups in the outer layer and their lateral pressure increase, which ultimately leads to a transition to large discs-like micelles. The thickness of discs is close to $2l$ throughout the entire D2 region, whereas their radius decreases with increasing $f$. Thick layers or large spheres are not formed by regular copolymer as for block copolymers, since then many $P$ groups would be in the condensed polymer. Thus, the block architecture of the copolymer, compared to the copolymer with regularly distributed amphiphilic monomer units, promotes the formation of micelles.

At a larger volume of polar group ($\upsilon_P/\upsilon_S = 2.5$), transitions from homogeneous to layered precipitate, and then to disc-shaped micelles, occur at smaller values of $f$ (compare Figs. 8a and 8b). Such shifts can be explained by a higher lateral pressure of bulky polar groups in the outer adjacent layer. Long macromolecules, consisting mostly of amphiphilic monomer units, can form also complex micelles consisting of beads (CM2). The shape of beads is disc-like near the transition line and becomes more spherical with increasing $f$, whereas the disc thickness is close to $2l$. Such transformations of micelles correspond to an increase in the volume $V_{\text{out}}$ of the adjacent layer, provided that the minimum characteristic size of micelles is fixed. These beads can consist of a single macromolecular segment or several segments. In the latter case, the beads become the branching points in bead sequences and can connect different macromolecules into a common aggregate.

The sizes of micelles are found to be practically independent of the chain length. The dependences of the sizes on the fraction $f$, shown in Fig. 7, would be the same for shorter chains. This is a consequence of the assumption of strong surface activity of amphiphilic monomer units. If the incompatible hydrophobic and polar moieties are strongly separated, then the sizes of the micelles are controlled by the volume ratio and sizes of these moieties, and not by the chain length. Micelles with an average sizes independent of chain length for copolymers of a given architecture with approximately the same composition were observed experimentally for hydrophilic poly(ethylene glycol) and hydrophobic dodecyl-graft copolymers in water.[10,11] The influence of the copolymer architecture on the average size of micelles and their weight distribution was analyzed. The alternative copolymer was found to form the smallest monodisperse particles in water relatively to random copolymer or copolymers with other grafted pendants distributions, with a necklace structure appearing for long enough polymers.[10]

In our consideration, the regular copolymer with uniformly distributed amphiphilic monomer units (composition 1:1) can be described as an alternative copolymer, and the regular copolymers are characterized by the same average sequence characteristics as random copolymers of the same composition, whereas the copolymer with a blocky architecture can be referred to as a copolymer with accumulated hydrophobic monomer segments (gradient and random block copolymers).[11] Larger particle sizes for the copolymers with blocky architecture agree with our predictions. The larger size can correspond to the formation of a particle core consisting of hydrophobic segments.

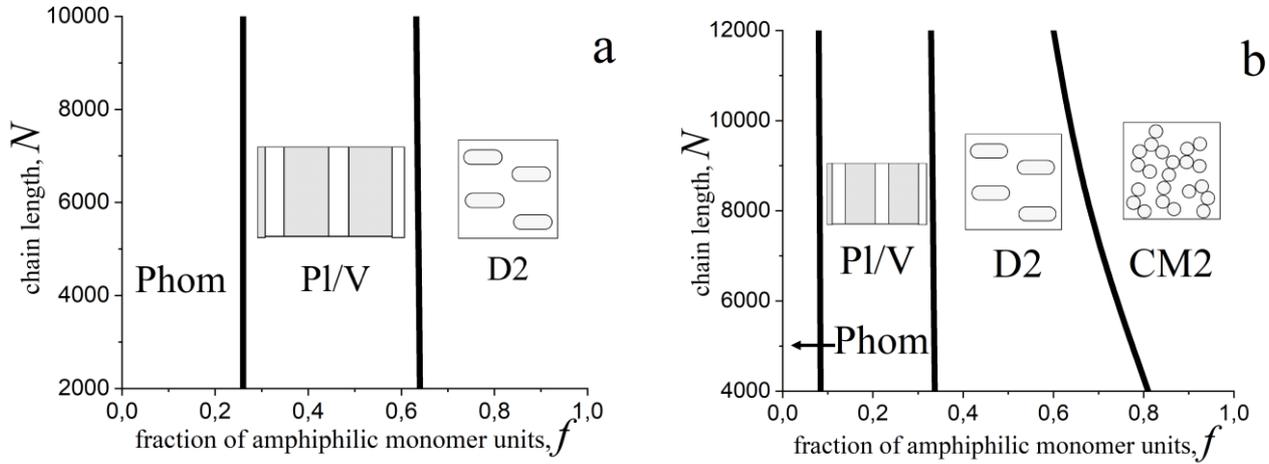

Figure 8. Morphological diagrams of regular copolymers in $f - N$ coordinates with (a) a moderate volume of the polar groups, $\upsilon_P/\upsilon_S = 1.5$, and (b) large volume, $\upsilon_P/\upsilon_S = 2.5$. Other parameters: $l = 3.75 \upsilon_S^{1/3}$, $\upsilon_H/\upsilon_S = a/\upsilon_S^{1/3} = 1.5$, $\gamma_0 = 0.8$, $\chi_{HP} = 4.4$, $\chi_{HS} = 4.2$.

For the block copolymers with large $P$ groups, the region of precipitate, which is layered, is very narrow in the diagrams (Fig. 9), i.e. only block copolymers with very long hydrophobic blocks can precipitate. The high lateral pressure of large $P$ groups in the outer adjacent layer leads to polymer micellization even for rather small fractions of amphiphilic blocks and promotes the spherical shape of the micelles. At $f > 0.1$ the block copolymer can form spherical particles with cores of hydrophobic blocks and additional small beads of amphiphilic blocks. Such beads provide maximum space for their side $P$ groups and they can be composed of segments belonging to different macromolecules, thus serving as links between particles. Therefore, granular complex micelles can be formed (CM1 and CM2). The beads are connected by hydrophobic backbone spacers and surrounded by adjacent layers containing polar groups. It is energetically favourable to shorten the spacers between beads, however the excluded volume interactions between the polar groups prevent overlapping of adjacent layers, so that the bead sequences can resemble granular filaments.

Changing the interaction forces between polar groups, i.e. the value of the Flory-Huggins parameter $\chi_{PS}$ ($\chi_{PS} = 0$ in our consideration), can influence the overlap of adjacent layers of the beads, ultimately leading to the formation of cylinder or worm-like micelles. For example, in computer simulations of micelles of amphiphilic homopolymers with polar main chains and hydrophobic pendants,[33] cylinder-like micelles were predicted only for negative values of $\chi_{PS}$.

With small fractions $f$ of amphiphilic blocks, particles with cores of hydrophobic blocks consist of a large number of macromolecules and have a large size. With increasing $f$, their aggregation number and radius decrease and, for short bond lengths $l$ (or very long macromolecules), their core consists of single hydrophobic blocks, whereas amphiphilic blocks form surrounding layers and additional beads (region CM1′ in Figs. 9). For longer lengths $l$ (or shorter macromolecules), there is an intermediate region of isolated spheres (IS), consisting of only a few macromolecules. These micelles can be stable in solution due to the repulsion between the surrounding P groups of different particles. Further increase of $f$ leads to homogeneous granular micelles (CM2) or, for long bond lengths or short macromolecules, to homogeneous isolated spherical micelles. At shorter bond lengths $l$, the beads and core-shell particles have smaller sizes and larger total surface areas, facilitating chains unfolding into coil conformations (Fig. 9b).

Single molecules of block copolymer with large $P$ groups also form a bead system (a necklace-like conformation) at large fractions $f$ of the amphiphilic block,[25] whereas they can form only one spherical particle with a core of a hydrophobic block under parameters corresponding to a precipitate or micelles with large core-shell particles in block copolymer solutions. The transition from a single particle to a necklace conformation depends significantly on the chain length $N$, i.e. larger chains begin to self-assemble into beads at smaller fractions $f$. Unlike a copolymer solution, single macromolecules form particles (globules) with an amphiphilic layer mostly thinner than $l$. Chains of different lengths form globules of different sizes, and the thickness $\Delta R$ of their amphiphilic layer is proportional to the globule radius $R$ at a fixed $f$. For longer chains this radius $R$ is larger and, hence, the thickness $\Delta R$ becomes equal to $l$ at smaller fractions $f$. The formation of a thicker layer of the amphiphilic block is unfavourable, and longer chains begin to form a necklace-like conformation.

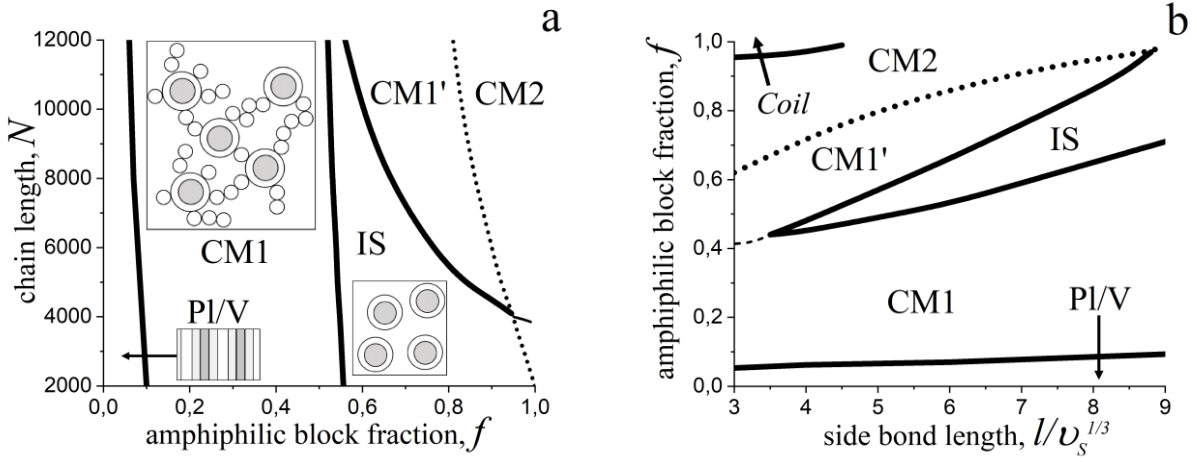

Figure 9. Morphological diagrams of block copolymers with a large volume of the polar groups, $\upsilon_P/\upsilon_S = 2.5$, in the coordinates (a) $f - N$ for $l = 6\upsilon_S^{1/3}$ and (b) $l - f$ for $N = 8000$. Other parameters: $\upsilon_H/\upsilon_S = a/\upsilon_S^{1/3} = 1.5$, $\gamma_0 = 0.8$, $\chi_{HP} = 4.4$, $\chi_{HS} = 4.2$.

Each of the figures 5-9 corresponds to a specific value of the volume $\upsilon_P$ of the $P$ group and their comparison permits to trace the role of $\upsilon_P$ and the length of the bond or chain separately. In practice, copolymer synthesis via grafting of polar side chains of different lengths produces copolymers with a volume of pendants proportional to their length. To monitor solution morphology of polymers with different lengths of the side chains, we present diagrams in the case of a linear relationship between the length $l$ and volume $\upsilon_P$ of the side group (Fig. 10). The movement from left to right in the diagram at a fixed $f$ corresponds to an increase in the length of the side chains at a fixed number of them. It is also worth noting that, in terms of the dimer model, an increase of the polar group volume $\upsilon_P$ can be related to an increase of the value of the second virial coefficient $B_{HP}$ describing the interactions between $H$ and $P$ groups: $B_{HP} = (\chi_{HP} - \chi_{HS} + \upsilon_H/\upsilon_S)\upsilon_P$.[26]

The effect of simultaneous changes in side chain length and volume on solution morphology is much more pronounced compared to the diagrams at the fixed $v_P$ (Figs. 5-9). For block copolymers (Fig. 10a), transitions from a precipitate with spherical cavities to a precipitate with cylindrical cavities, then to a lamellar precipitate, and to micelles occur at smaller $f$ with increasing $l$ along with $v_P$. If the thickness of the amphiphilic block layer is equal to $l$ as well as the cavity size, then the total volume of cavities $V_{\text{out}}$ is almost does not depend on $l$ under a fixed $f$. At the same time, the volume fraction of $P$ groups in the cavities, $\psi_P$, is proportional to $v_P$ (and, correspondingly, $l$) and, then, the layer free energy $F_{\text{layer}}$ (Eq. (9)) and pressure in this layer are also very sensitive to $l$. As a result, these transitions require much fewer $P$ groups (a smaller fraction $f$) for larger $l$.

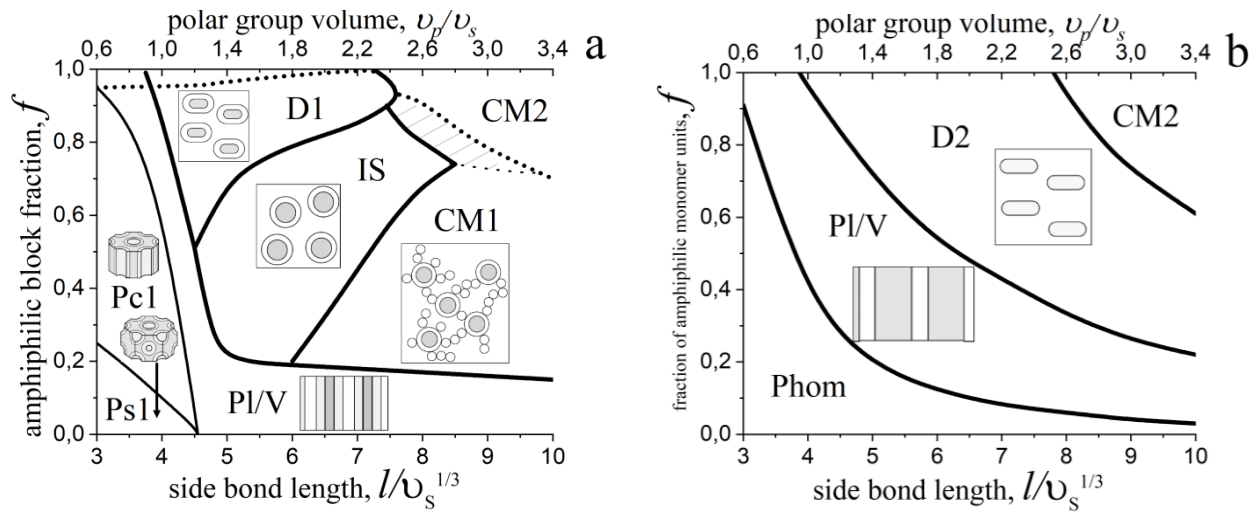

Figure 10. Morphological diagram of (a) block copolymers and (b) regular copolymers in the coordinates $l$ (or $v_P/v_S$) and $f$ at $N = 8000$, $v_H/v_S = a/v_S^{1/3} = 1.5$, $\gamma_0 = 0.8$, $\chi_{HP} = 4.4$, $\chi_{HS} = 4.2$. The length $l$ is set to increase linearly with the volume of $P$ groups: $l/v_S^{1/3} = 2.5\, v_P/v_S + 1.5$. The area above the upper dotted line (a) corresponds to homogeneous lamellas or micelles. The shaded area describes complex micelles with cores consisting of single hydrophobic blocks.

Thin layers (at large $f$) transform with increasing $l$ into discs (D1) of approximately the same thickness, but with additional space for $P$ groups at the edges. Layers of moderate thickness (at moderate values of $f$) transform into spherical particles (IS) with a core of hydrophobic blocks, and thick layers (at small $f$) retain their shape with increasing $l$ along with $\upsilon_P$. The radius of the disc in the vicinity of the transition line is quite large; with increasing $l$ it decreases and discs transform into spherical particles. This transition proceeds at larger $f$ with increasing $l$ (and $\upsilon_P$), since the curvature effect is more pronounced for smaller micelles, reducing the volume fraction of $P$ groups in the outer layer, and hence a larger volume of $P$ groups is necessary for the transition.

With increasing $l$ and $\upsilon_P$ at a fixed $f$, the thickness of the amphiphilic layer and the radius of the spherical particles increase until the volume fraction $\psi_P$ becomes large enough ($\psi_P > 0.5$), i.e. the outer layer is nearly saturated by the polar groups. Further increase in $l$ and $\upsilon_P$ leads to a slight decrease in thickness of the amphiphilic layer and particle radius, whereas the pressure in the outer adjacent layer continue to grow. Finally, these particles transform into larger ones with a thinner layer of amphiphilic blocks accompanied by additional small particles consisting purely of segments of the amphiphilic blocks. This transition also proceeds at larger $f$ (for smaller micelles) with increasing $l$, since a larger volume of $P$ groups is necessary to create the same pressure in the outer layer of the particles with higher curvature. In the shaded area, the cores of particles are composed of single hydrophobic blocks (CM1′). For shorter macromolecules, this area is disappeared.

For copolymers with a uniform distribution of amphiphilic monomer units, the morphological diagram is close to symmetrical (Fig. 10b). Both an increase in the fraction $f$ and an increase in the length $l$ (along with $\upsilon_P$) leads to the transitions: homogeneous precipitate (Phom) – layers (Pl/V) – discs (D2) – complex micelles composed of beads (CM2). The total volume of polar groups in the system is proportional to the product $f\upsilon_P$. If it is small, then the contribution of unfavourable interactions between $H$ and $P$ groups in the homogeneous precipitate is less than the free energy required for the surface formation of layers. As the product $f\upsilon_P$ increases, the volume fraction of polar groups in the precipitate, $\varphi_P$, increases, and it becomes layered with the layers of thickness $2l$ and almost all $P$ groups outside. As long as the value of the volume fraction of $P$ groups in the outer layer, $\psi_P$, remains moderate, the total volume of the outer adjacent layers of the precipitate layers remains practically unchanged with $l$, since the total surface area is inversely proportional to the layer thickness, whereas the thickness of the outer layer is equal to $l$. Then $\psi_P$ is proportional to $f\upsilon_P$, and the steric interactions of the $P$ groups in the outer layer become more and more important. This leads to the formation of disc-like micelles, the radius of which decreases with increasing $f\upsilon_P$. The disc thickness grows with $l$ until the outer layer becomes almost saturated with the $P$ groups. Then, the thickness stops increasing and finally the discs transform into a system of interconnected beads (CM2).

In the previous theoretical consideration of micelles in solution of regular copolymers with pendant polar groups,[34] different structures were predicted depending on the surface tension of the flat surface $\gamma$ of the condensed polymer and the bending moduli. With a positive value of the spontaneous bending modulus $\kappa_1$, a macrophase (precipitate), spherical or cylindrical mesoglobules, or a gyroid structure were found at positive $\gamma$, and layers (vesicles) at negative $\gamma$. For a negative value of $\kappa_1$, a macrophase, inverse spheres, cylinders and gyroid (precipitates with cavities) were predicted at positive $\gamma$, and also layers (vesicles) at negative $\gamma$.

Intuitively, our consideration in the limit of very high surface activity of H-P dimers can be related to a negative surface tension $\gamma$ that corresponds to a layered structure in Ref. 34. However, our model (Eqs. (1)-(9)) cannot be described by the free energy in the form $F = F_{\text{bulk}} + \gamma_{\text{flat}}S + F_s$, where the bulk free energy $F_{\text{bulk}} = const$, the surface tension $\gamma_{\text{flat}} = const$, and the morphology is controlled only by the bending moduli. Besides, the characteristic size of the micelles of regular copolymers in our consideration appears to be comparable with the side bond length $l$ contrary to the assumption[34] of large sizes.

Summarising the analysis of morphologies in relation to the macromolecular architecture (Figs. 5-10), we observe that both block and regular copolymers self-assemble into systems of spherical particles in the case of sufficiently large polar pendants and high fractions of amphiphilic monomer units. In particular, necklace-like micelles and branched granular aggregates are formed, where small particles (beads) of condensed amphiphilic block segments of different chains play the role of stickers, ensuring the connectivity of the entire aggregate. Such structures were observed in experiments for solutions of locally amphiphilic graft polymers,[4,6,7] for amphiphilic random copolymers,[9] and block polymers containing locally amphiphilic blocks.[16-21]

Polymers with smaller polar pendants are predicted to form other morphologies, such as disc-like micelles or layers/vesicles. These structures were also observed experimentally in solutions of the amphiphilic graft copolymers,[5] and, in addition, many examples of vesicle-forming systems can be found in the review.[2] Condensed polymer morphologies with nanometer-sized heterogeneities can be detected, for example, using transmission electron microscopy, as for the "large compound vesicles"[12] with hydrophilic cavities formed by locally amphiphilic homopolymers. The structure of micelles[12] depended on polymer lengths, and a branched cylinder system was observed for mixed long and short polymers. In our consideration, chain lengths influence very slightly the morphology diagrams due to strong repulsion between hydrophobic and polar moieties under the condition of high chain flexibility. The noticeable influence of chain lengths on the structural characteristics[12] can be related to the dependence of conformational energy of chain packing in micelles on the presence of chain ends in the case of a moderate level of interaction energies.

## 4. Conclusions

Summarizing, we have analyzed the relation between the architecture and composition of the macromolecules and the solution morphology for locally amphiphilic polymers. Using the mean-field theory, the favorable structure types and geometric parameters of the micelles in solution or cavities in a condensed polymer have been predicted. In the case of high surface activity of locally amphiphilic chain segments, the characteristic size of the domains of amphiphilic blocks or chains is of the order of the average distance between adjacent hydrophobic and polar moieties of the chain. In the case of sufficiently large polar groups, a necklace conformation of amphiphilic chains is favorable. The core-shell particles and beads of the amphiphilic blocks can be composed of distant segments of the same chain or segments of different chains, thus being branching points in bead sequences. Thus, we propose to consider the condensed locally amphiphilic polymers as intramicellar links that provide both the connectivity of polymer chain aggregates and the local heterogeneity of their structure. This explains why the granular branched micelles are observed in many experimental studies of locally amphiphilic polymers. The blockiness in the arrangement of amphiphilic monomer units is shown to promote the formation of spherical particles with cores of hydrophobic blocks. Although the size of these particles and beads is almost independent of polymer length, longer polymers should facilitate the formation of larger aggregates and accelerate the aggregation process.

For moderate sizes of polar pendants, their volume, the chain composition and distribution pattern of locally amphiphilic segments control self-assembly into other morphologies such as disc-like micelles, vesicles or layers, as well as inverse structures of large aggregates or precipitates. These structures are less frequently observed, however they also hold promise for testing as drug carriers, elements of artificial cells or building blocks for tissue engineering.

As we describe the micelles with clearly segregated domains of hydrophobic blocks, amphiphilic blocks and polar shell in the limit of high surface activity of amphiphilic monomer units and strong hydrophobicity of the backbone, the transition conditions for polymer micellization from the unfolded state cannot be exactly determined. However, it was possible to analyze the influence of the interaction parameters on the micelle shapes and morphological transitions and to predict a shift in transformations from precipitates to vesicles/layers and further to micelles, to smaller fractions of amphiphilic blocks with increase in the solvent quality or surface activity of amphiphilic monomer units.

Always, the aggregation of locally amphiphilic polymers is characterized by the high surface-to-volume ratio that makes the micelles sensitive to even small concentration of surface reagents. We hope that our investigation reveals peculiarities of this class of polymers and it can be useful for the analysis of polymer micellization for many practical purposes.

## Acknowledgments

E. N. G. is thankful to the French National program Pause for the financial support.

## Conflicts of interest

There are no conflicts to declare.

## Data availability statement

A method for calculating the micelle morphologies other that isolated spheres is presented in *Supporting information*. The computer code and the data generated are available from the authors upon reasonable request.

# Aggregation of hydrophobic-amphiphilic block copolymers

Sophia A Pavlenko[1], Elena N Govorun[2*]

[1] *Faculty of Physics, Lomonosov Moscow State University, 119991 Moscow, Russia*

[2] *UMR CNRS 7083 Gulliver, ESPCI Paris, PSL Research University, 75005 Paris, France*
*elena.govorun@espci.fr, elenagovorun653@gmail.com*

We study the solution morphology of block copolymers with amphiphilic and hydrophobic blocks. The amphiphilic monomer units are represented as dimers with hydrophobic groups in the main chain and side polar groups. The macromolecules can form micelles of different shapes or precipitate (Figs. 3, 4 *in the main text*), and to construct morphological diagrams, calculations of the free energies of all structures are necessary. The geometrical parameters and the free energy terms are described in the main text for isolated spheres (IS). Below, the other structure types of micelles and precipitates are described.

Hydrophobic blocks consist of $N_H$ hydrophobic monomer units of volume $v_H$ and amphiphilic blocks consist of $N_A$ dimer monomer units with hydrophobic groups of the same volume and polar groups of volume $v_P$. The volume of solvent molecule is $v_S$. The length of the macromolecule backbone is $N = N_H + N_A$ and the amphiphilic block fraction $f = N_A/N$. The monomer unit size along the backbone chain is equal to *a*, the length of bond between hydrophobic and polar groups in the amphiphilic monomer units is *l*. The pairwise interactions of hydrophobic groups with solvent and with polar groups are characterized by the Flory-Huggins parameters $\chi_{HS}$ and $\chi_{HP}$, respectively. For the interactions between polar groups and solvent, the zero value $\chi_{PS} = 0$ is taken.

## 1. Smoothed disc-like and cylindrical micelles with a hydrophobic core

It is assumed that a micelle of volume $V$ consists of a homogeneous core of volume $V_0$ and a homogeneous shell of constant thickness $\Delta R$. The core and the shell are composed of hydrophobic and amphiphilic blocks, respectively. The surface area of the micelle is $S$, the surface area of the core (or the area of the core-shell interface) is $S_0$. The number of macromolecules forming the micelle (the aggregation number) is $m$, the volume fractions of hydrophobic groups in the core and in the shell are $\varphi_{H0}$ and $\varphi_H$, respectively. They are determined by the relations (the same as Eq. (1) in the main text):

$$\varphi_{H0} = \frac{mN_H \upsilon_H}{V_0}, \quad \varphi_H = \frac{mN_A \upsilon_H}{V - V_0}. \tag{S1}$$

From the incompressibility condition, the solvent volume fraction in the core is $\varphi_{S0} = 1 - \varphi_{H0}$.

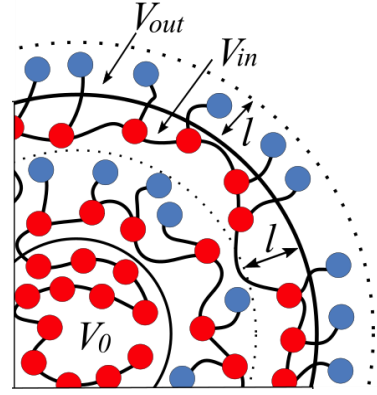

The hydrophobic groups of amphiphilic monomer units in the surface layer of the micelle of thickness $l$ and volume $V_{in}$ have their polar groups outside in the outer adjacent layer of thickness $l$ and volume $V_{out}$ (see the schematic figure). Let $N_{Pout}$ be the number of such polar groups. If the thickness of the amphiphilic shell $\Delta R > l$, then $N_{Pout} = \varphi_H V_{in}/\upsilon_H$. Otherwise, all polar groups are in the outer adjacent layer: $N_{Pout} = mN_A$.

The volume fractions of polar groups in the shell and in this layer are

$$\varphi_P = \frac{(mN_A - N_{Pout})\upsilon_P}{V - V_0}, \quad \psi_P = \frac{N_{Pout}\upsilon_P}{V_{out}}. \tag{S2}$$

The solvent volume fraction in the shell is $\varphi_S = 1 - \varphi_H - \varphi_P$.

The disc-like micelle has the thickness $2R$ and the circles of radius $r$ at the bases, the thickness of the core is $2R_0$ (Fig. S1a). The cylinder-like micelle has the radius $R$, the height $L$, and the core of radius $R_0$ (Fig. S1b). The micelle shapes are smoothed so that, in the cross section, the side surface of the disc-like micelle and the top and base surfaces of the cylinder-like micelle are semicircles of radius $R$, and the similar semicircles in the cross-sections of the core are of radius $R_0$. These shapes transform into a spherical particle with a spherical core if $r \to 0$ for a disc-like micelle and if $L \to 0$ for a cylinder-like micelle.

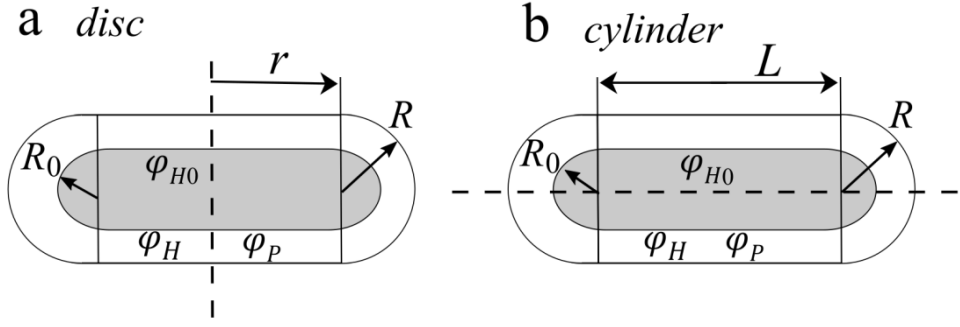

Figure S1. Cross sections of the smoothed (a) disc-like micelle and (b) cylinder-like micelle with the core of hydrophobic blocks and with the shell of constant thickness. The dashed lines are the rotation axes of the shapes.

The free energy of these core-shell micelles consists of the same terms as the free energy of spherical micelles (Eq. (3)):

$$F_{c-s} = F_{bulk0} + F_{bulk} + F_{inter} + F_{surf} + F_{layer}, \quad (S3)$$

where all contributions are as in Eqs. (4)-(9). The bulk free energy of the core

$$\frac{F_{bulk0}}{mNk_BT} = (1-f)\left(\chi_{HS}\varphi_{S0} + \frac{\upsilon_H}{\upsilon_S}\frac{\varphi_{S0}}{\varphi_{H0}}\ln\varphi_{S0}\right), \quad (S4)$$

the shell

$$\frac{F_{bulk}}{mNk_BT} = f\left(\chi_{HP}\varphi_P + \chi_{HS}\varphi_S + \frac{\upsilon_H}{\upsilon_S}\frac{\varphi_S}{\varphi_H}\ln\varphi_S\right). \quad (S5)$$

The free energy of a core-shell interface

$$\frac{F_{inter}}{mNk_BT} = \frac{S_0}{mN\upsilon_S^{2/3}}(\varphi_{H0}-\varphi_H)\left(\gamma_0 + \frac{1}{2}\left((\chi_{HP}-\chi_{HS})\varphi_P + \chi_{HS}(\varphi_{H0}-\varphi_H)\right)\right) \quad (S6)$$

and the surface free energy

$$\frac{F_{surf}}{mNk_BT} = \frac{S}{mN\upsilon_S^{2/3}}\varphi_H\left(\gamma_0 + \frac{1}{2}\left((\chi_{HP}-\chi_{HS})(\psi_P-\varphi_P) + \chi_{HS}\varphi_H\right)\right). \quad (S7)$$

The free energy of the outer adjacent layer

$$\frac{F_{layer}}{mNk_BT} = \frac{V_{out}}{mN\upsilon_S}(1-\psi_P)\ln(1-\psi_P) \quad (S8)$$

takes into account only the translation entropy of the solvent molecules. It is assumed that the core size and the shell thickness exceed the mean diameter of hydrophobic monomer unit: $R_0 > d_H$, $\Delta R = R - R_0 > d_H$ ($d_H = (6\upsilon_H/\pi)^{1/3}$). If these conditions cannot be fulfilled, only homogeneous micelles are considered.

The disc-like micelle can be characterized by the aspect ratio $\theta = r/R$ and the radius $R$ instead of the radii $r$ and $R$. The aspect ratio of the core is $\theta_0 = r/R_0$. All volumes and areas in Eqs. (S1)-(S8) can be determined using Pappus's centroid theorem:

$$\text{Disc:} \quad V = \pi R^3 \left(2\theta^2 + \pi\theta + \frac{4}{3}\right), \quad S = 2\pi R^2(\theta^2 + \pi\theta + 2), \tag{S9}$$

$$V_{\text{out}} = 2\pi l \left(\theta^2 R^2 + \left(R + \frac{l}{2}\right)\left(\pi\theta R + 4\frac{R^2 + lR + l^2/3}{2R + l}\right)\right),$$

$$V_{\text{in}} = 2\pi l \left(\theta^2 R^2 + \left(R - \frac{l}{2}\right)\left(\pi\theta R + 4\frac{R^2 - lR + l^2/3}{2R - l}\right)\right),$$

$$V_0 = R_0^3 \left(2\pi\theta_0^2 + \pi^2\theta_0 + \frac{4}{3}\pi\right), \quad S_0 = 2\pi R_0^2(\theta_0^2 + \pi\theta_0 + 2).$$

For a cylindrical micelle, the aspect ratios are $\theta_c = L/R$ and $\theta_{0c} = L/R_0$. The volumes and areas are

$$\text{Cylinder:} \quad V = \pi R^3 \left(\theta_c + \frac{4}{3}\right), \quad S = 2\pi R^2(\theta_c + 2), \tag{S10}$$

$$V_{\text{out}} = \pi l\left(4(R^2 + lR + l^2/3) + \theta_c R(2R + l)\right),$$

$$V_{\text{in}} = \pi l\left(4(R^2 - lR + l^2/3) + \theta_c R(2R - l)\right),$$

$$V_0 = \pi R_0^3 \left(\theta_{0c} + \frac{4}{3}\right), \quad S_0 = 2\pi R_0^2(\theta_{0c} + 2).$$

For the disc and cylinder-like micelles (Fig. S1), the free energy $F$ (S3) can be represented as a function of four independent variables rather than three variables for the spherical core-shell micelles. By minimizing the free energy $F$ with respect to these variables, the equilibrium values of the micelle sizes ($r$, $R$, $R_0$ for a smoothed disc and $L$, $R$, $R_0$ for a smoothed cylinder), the aggregation number $m$ and the free energy of these morphologies are calculated.

## 2. Vesicle-like micelles

Let micelles, composed of $m$ macromolecules, be spherical shells (vesicles) of outer radius $R$ and inner radius $r$. In the shell, the hydrophobic blocks form a separate spherical layer (a hydrophobic core) with outer radius $R_0$, inner radius $R_1$, volume $V_0$, and thickness $\Delta R_0 = R_0 - R_1$ (Fig. S2). The amphiphilic blocks form an inner spherical layer of volume $V_1$ and an outer layer of volume $V_2$.

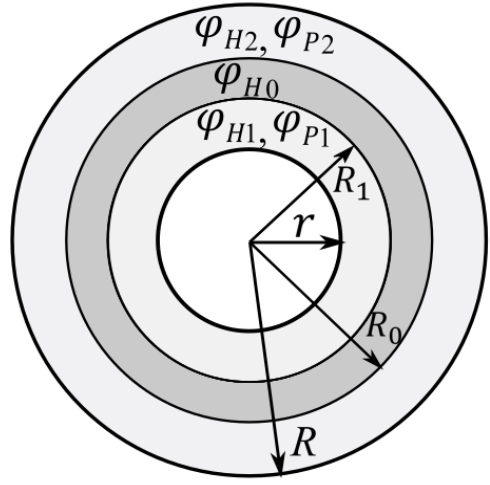

Figure S2. Scheme of a vesicle-like micelle with a core of hydrophobic blocks in the shape of a spherical layer (dark gray) and two layers of amphiphilic blocks (light gray).

Let $\varphi_{H0}$ be the volume fraction of hydrophobic groups in the core, $\theta_v$ be the fraction of amphiphilic monomer units in the outer layer respectively to the total number of amphiphilic monomer units, $\varphi_{H1}$ and $\varphi_{H2}$ be the volume fractions of hydrophobic groups in the inner and outer amphiphilic layers, respectively. Then, the solvent volume fraction in the core is $\varphi_{S0} = 1 - \varphi_{H0}$ and the fraction of amphiphilic monomer units in the inner layer is $1 - \theta_v$. The volume fractions of polar groups in the inner and outer layers of the shell are denoted by $\varphi_{P1}$ and $\varphi_{P2}$. Then, the solvent volume fractions in the inner and outer layers are $\varphi_{S1} = 1 - \varphi_{H1} - \varphi_{P1}$ and $\varphi_{S2} = 1 - \varphi_{H2} - \varphi_{P2}$. The volume fractions of polar groups in the layers adjacent to the shell inside and outside the vesicle are denoted by $\psi_{P1}$ and $\psi_{P2}$.

The total volume of the shell is

$$V = V_0 + V_1 + V_2 = \frac{4\pi}{3}(R^3 - r^3)$$

where the volume of the hydrophobic core is

$$V_0 = \frac{4\pi}{3}(R_0^3 - R_1^3),$$

and the volumes of the amphiphilic layers are

$$V_1 = \frac{4\pi}{3}(R_1^3 - r^3), \quad V_2 = \frac{4\pi}{3}(R^3 - R_0^3).$$

The outer surface area of the micelle is $S = 4\pi R^2$, the surface area of the cavity is $S_{cav} = 4\pi r^2$, and the inner and outer surface areas of the hydrophobic core are $S_{0(in)} = 4\pi R_1^2$ and $S_{0(out)} = 4\pi R_0^2$.

The volume fractions of hydrophobic groups in the core and in the amphiphilic layers are

$$\varphi_{H0} = \frac{mN_H \upsilon_H}{V_0}, \quad \varphi_{H1} = \frac{(1-\theta_v)mN_A \upsilon_H}{V_2}, \quad \varphi_{H2} = \frac{\theta_v mN_A \upsilon_H}{V_1}. \quad (S11)$$

If the thickness of the inner layer of amphiphilic blocks $R_1 - r < l$, then all polar groups of monomer units of this layer are in the cavity and its number is $N_{Pout1} = (1-\theta_v)mN_A$. If $R_1 - r > l$, then only part of these groups is in the cavity and their number is equal to that of hydrophobic groups in the spherical layer of thickness $l$ adjacent to the cavity: $N_{Pout1} = \varphi_{H1} V_{in1}/\upsilon_H$, where $V_{in1} = 4\pi l(r^2 + rl + l^2/3)$ is the volume of this inner adjacent layer. Then, the volume fractions of polar groups are

$$\varphi_{P1} = \frac{((1-\theta_v)mN_A - N_{Pout1})\upsilon_H}{V_1}, \quad \psi_{P1} = \frac{N_{Pout1}}{V_{out1}}, \quad (S12)$$

where $V_{out1} = 4\pi l(r^2 - rl + l^2/3)$ if the cavity radius $r > l$, and $V_{out1} = 4\pi r^3/3$ if $l/2 < r < l$. The vesicles with the cavities of smaller size are not considered.

For the outer surface of the vesicle, similarly, the number of polar groups outside is $N_{Pout2} = \theta_v mN_A$, if the thickness of the outer shell layer of amphiphilic blocks $R - R_0 < l$. For $R - R_0 > l$, the number $N_{Pout2} = \varphi_{H2} V_{in2}/\upsilon_H$, where $V_{in2} = 4\pi l(R^2 - Rl + l^2/3)$. Then, the volume fractions are

$$\varphi_{P2} = \frac{(\theta_v mN_A - N_{Pout1})\upsilon_H}{V_2}, \quad \psi_{P2} = \frac{N_{Pout2}\upsilon_H}{V_{out2}}. \quad (S13)$$

where the volume of the outer adjacent layer of thickness $l$ is $V_{out2} = 4\pi l(R^2 + Rl + l^2/3)$.

The free energy of vesicles consists of the same terms as the free energy of simple core-shell micelles (Eqs. (3) and (S3)):

$$F_v = F_{bulk0} + F_{bulk} + F_{inter} + F_{surf} + F_{layer}$$

with the core free energy $F_{bulk0}$ described by Eq. (S4). All other contributions consist of two parts describing the inner and outer layers of the vesicle:

$$F_{bulk} = F_{bulk(in)} + F_{bulk(out)}, \quad F_{inter} = F_{inter(in)} + F_{inter(out)},$$
$$F_{surf} = F_{surf(in)} + F_{surf(out)}, \quad F_{layer} = F_{layer(in)} + F_{layer(out)},$$

which are given by

$$\frac{F_{bulk(in)}}{mNk_BT} = (1-\theta_v)f\left(\chi_{HP}\varphi_{P1} + \chi_{HS}\varphi_{S1} + \frac{\upsilon_H}{\upsilon_S}\frac{\varphi_{S1}}{\varphi_{H1}}\ln\varphi_{S1}\right), \quad (S14)$$

$$\frac{F_{bulk(out)}}{mNk_BT} = \theta_v f\left(\chi_{HP}\varphi_{P2} + \chi_{HS}\varphi_{S2} + \frac{\upsilon_H}{\upsilon_S}\frac{\varphi_{S2}}{\varphi_{H2}}\ln\varphi_{S2}\right)$$

for the bulk free energies of the layers,

$$\frac{F_{inter(in)}}{mNk_BT} = \frac{S_{0(in)}}{mN\upsilon_S^{2/3}}(\varphi_{H0} - \varphi_{H1})\left(\gamma_0 + \frac{1}{2}\left((\chi_{HP} - \chi_{HS})\varphi_{P1} + \chi_{HS}(\varphi_{H0} - \varphi_{H1})\right)\right), \quad (S15)$$

$$\frac{F_{\text{inter(out)}}}{mNk_BT} = \frac{S_{0(\text{out})}}{mN\upsilon_S^{2/3}}(\varphi_{H0} - \varphi_{H2})\left(\gamma_0 + \frac{1}{2}\bigl((\chi_{HP} - \chi_{HS})\varphi_{P2} + \chi_{HS}(\varphi_{H0} - \varphi_{H2})\bigr)\right)$$

for the free energies of the inner and outer interfaces between the core and the amphiphilic layers,

$$\frac{F_{\text{surf(in)}}}{mNk_BT} = \frac{S_{\text{cav}}}{mN\upsilon_S^{2/3}}\varphi_{H1}\left(\gamma_0 + \frac{1}{2}\bigl((\chi_{HP} - \chi_{HS})(\psi_{P1} - \varphi_{P1}) + \chi_{HS}\varphi_{H1}\bigr)\right), \quad \text{(S16)}$$

$$\frac{F_{\text{surf(out)}}}{mNk_BT} = \frac{S}{mN\upsilon_S^{2/3}}\varphi_{H2}\left(\gamma_0 + \frac{1}{2}\bigl((\chi_{HP} - \chi_{HS})(\psi_{P2} - \varphi_{P2}) + \chi_{HS}\varphi_{H2}\bigr)\right)$$

for the free energies at the inner and outer surfaces, and

$$\frac{F_{\text{layer(in)}}}{mNk_BT} = \frac{V_{\text{out1}}}{mN\upsilon_S}(1 - \psi_{P1})\ln(1 - \psi_{P1}), \quad \text{(S17)}$$

$$\frac{F_{\text{layer(out)}}}{mNk_BT} = \frac{V_{\text{out2}}}{mN\upsilon_S}(1 - \psi_{P2})\ln(1 - \psi_{P2})$$

for the bulk free energies of inner and outer adjacent layers.

Thus, the free energy of vesicles $F_v$ can be represented as a function of six independent variables. By minimizing the free energy $F_v$ with respect to these variables, the equilibrium values of the vesicle sizes, the volume fractions, and the aggregation number $m$ can be found together with the free energy of this morphology.

## *3. Complex micelles of spherical particles*

Complex micelles contain core-shell spherical particles and smaller particles (beads) of amphiphilic monomer units (Fig. S3). The latter particles can be composed of blocks belonging to different polymer chains and thus be the links or branching points in sequences of particles. Then, such polymer aggregates in solution can be characterized as complex or compound micelles with the branching granular structure.

Let the core-shell particles in the complex micelles be described by the same parameters as the single core-shell spherical particles (Model section *in the main text*): the radii $R_0$ and $R$, the volumes $V_0$ and $V$, the volume fractions of hydrophobic groups $\varphi_{H0}$ and $\varphi_H$, and the volume fractions of polar groups $\varphi_P$ and $\psi_P$. Let $m$ be the number of macromolecules per core-shell particle in a complex micelle, $M$ be the number of beads per core-shell particle in the complex micelle, $q$ be the fraction of amphiphilic monomer units forming the shells, and $1 - q$ forming the homogeneous beads.

Our preliminary analysis has demonstrated that if the appearance of an additional bead is energetically favourable, then the bead radius is equal to $R_b = l$ so that all polar groups are repelled from the bead. Then, the bead volume $V_{\text{beed}} = 4\pi l^3/3$. Let $\varphi_{H1}$ be the volume fraction of hydrophobic groups in such beads and $\psi_{P1}$ be the volume fraction of polar groups in its outer adjacent layer of volume $V_{out1} = 28\pi l^3/3$.

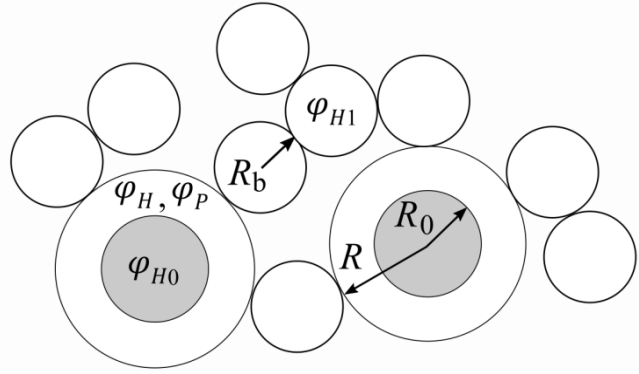

Figure S3. Scheme of a part of a complex micelle with the core-shell particles and beads of amphiphilic blocks. Hydrophobic blocks form the cores, amphiphilic blocks form the shells and additional beads. Such beads can link the core-shell particles and can also serve as branching points in bead sequences.

For the core-shell spherical particle, the volume fractions are equal to

$$\varphi_{H0} = \frac{mN_H \upsilon_H}{V_0}, \quad \varphi_H = \frac{mqN_A \upsilon_H}{V - V_0}, \quad \varphi_P = \frac{(mqN_A - N_{Pout})\upsilon_P}{V - V_0}, \quad \psi_P = \frac{N_{Pout}\upsilon_P}{V_{out}}. \quad (S18)$$

where $N_{Pout} = mN_A$ if $R - R_0 < l$ and $N_{Pout} = \varphi_H V_{in}/\upsilon_H$ if $R - R_0 \geq l$, where the volume of the inner adjacent layer is $V_{in} = 4\pi l(R^2 - Rl + l^2/3)$. The volume fractions of hydrophobic groups in the beads and polar groups in the outer adjacent layers of the beads are

$$\varphi_{H1} = \frac{m(1-q)fN_A\upsilon_H}{MV_{beed}}, \quad \psi_{H1} = \frac{m(1-q)fN_A\upsilon_P}{MV_{out1}}. \quad (S19)$$

The free energy $F_{c.m.}$ of the complex micelle per core-shell particle consists of the free energy $F_{c-s}$ of the core-shell particle (Eq. (S3)) and the free energy of the beads, $F_{beeds}$:

$$F_{c.m.} = F_{c-s} + F_{beeds}. \quad (S20)$$

For a core-shell particle, the contribution $F_{bulk0}$ to the free energy $F_{c-s}$ is given by Eq. (S4), the bulk free energy of the shell is

$$\frac{F_{bulk}}{mNk_BT} = qf\left(\chi_{HP}\varphi_P + \chi_{HS}\varphi_S + \frac{\upsilon_H}{\upsilon_S}\frac{\varphi_S}{\varphi_H}\ln\varphi_S\right), \quad (S21)$$

and the contributions $F_{inter}$, $F_{surf}$, and $F_{layer}$ are given by Eqs. (S6)-(S8). For the beads, the contributions to the free energy can be written similarly to those for $F_{c-s}$ but without considering the cores:

$$F_{beeds} = F_{bulk(b)} + F_{surf(b)} + F_{layer(b)}, \quad (S22)$$

where

$$\frac{F_{\text{bulk(b)}}}{mNk_BT} = (1-q)f\left(\chi_{HS}(1-\varphi_{H1}) + \frac{v_H}{v_S}\frac{1-\varphi_{H1}}{\varphi_{H1}}\ln(1-\varphi_{H1})\right), \quad (S23)$$

$$\frac{F_{\text{surf(b)}}}{mNk_BT} = \frac{MS_{\text{beed}}}{mNv_S^{2/3}}\varphi_{H1}\left(\gamma_0 + \frac{1}{2}\left((\chi_{HP}-\chi_{HS})\psi_{P1}+\chi_{HS}\varphi_{H1}\right)\right), \quad (S24)$$

where $S_{\text{beed}} = 4\pi l^2$, and

$$\frac{F_{\text{layer(b)}}}{mNk_BT} = \frac{MV_{\text{out1}}}{mNv_S}(1-\psi_{P1})\ln(1-\psi_{P1}). \quad (S25)$$

The free energy $F_{\text{c.m.}}$ (S20) is minimized with respect to the radii $R_0$ and $R$, the fraction $q$, and the numbers $m$ and $M$. The equilibrium values of these parameters characterize the structure of complex micelles. The corresponding value of the free energy per monomer unit is compared with those of the other structures.

## *4. Precipitate*

Polymer with long hydrophobic blocks and copolymers with a small fraction of regularly distributed amphiphilic monomer units can precipitate or form macroscopic agglomerates in solution. We consider a homogeneous precipitate and inhomogeneous precipitates with domains containing only polar groups and solvent. For the homogeneous precipitate, the free energy per monomer unit is determined only by the bulk contribution:

$$\tilde{F}_{\text{pr0}} = \tilde{F}_{\text{bulk}} = k_BT\left(\chi_{HP}\varphi_P + \chi_{HS}\varphi_S + \frac{v_H}{v_S}\frac{\varphi_S}{\varphi_H}\ln\varphi_S\right), \quad (S26)$$

where $\varphi_H$ and $\varphi_P$ are the volume fraction of hydrophobic and polar groups, respectively, $\varphi_S = 1 - \varphi_H - \varphi_P$ is the solvent volume fraction. By minimizing the free energy $\tilde{F}_{\text{pr0}}$ with respect to $\varphi_H$ and $\varphi_P$, their equilibrium values can be found.

The inhomogeneous precipitates considered are a layered structure with infinite layers and precipitates with cavities, where hydrophobic groups form a continuous phase with infinite cylindrical or spherical pores. In Fig. S4, the schemes of layered morphologies are presented for homogeneous lamellas formed by the regular copolymer with a homogeneous distribution of amphiphilic monomer units and for lamellas with middle layers of hydrophobic blocks formed by the block copolymer.

The lamella thickness is denoted by $d$ and the distance between neighbour lamellas by $d_{\text{out}}$. The volume fraction of hydrophobic groups in the middle layer of thickness $d_0$ is denoted by $\varphi_{H0}$, the volume fractions of hydrophobic and polar groups in the side layers of amphiphilic blocks and in homogeneous lamellas are denoted by $\varphi_H$ and $\varphi_P$, the volume fraction of polar groups in the adjacent layers outside the lamellas by $\psi_P$. The total volume of the lamellas $V = S_l d$, where $S_l$ is their total area, and then their total surface area is equal to $S_{lam} = 2S = 2V/d$.

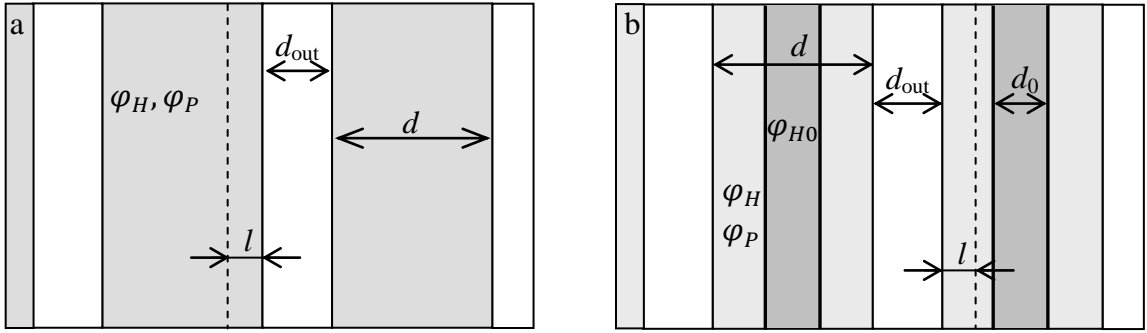

Figure S4. Scheme of a layered precipitate with (a) homogeneous polymer lamellas (Pl0) and (b) polymer lamellas with middle layers of hydrophobic blocks and side layers of amphiphilic blocks (Pl).

The free energy (per monomer unit) of homogeneous lamellas (Fig. S4a) contains the same contributions as those of homogeneous micelles of any shape:

$$\tilde{F}_{Pl0} = \tilde{F}_{bulk} + \tilde{F}_{surf} + \tilde{F}_{layer}, \tag{S27}$$

where the bulk free energy $\tilde{F}_{bulk}$ is given by Eq. (S26), the surface free energy per monomer unit is given by

$$\tilde{F}_{surf} = k_B T \frac{2\varphi_H}{\rho_S N v_S^{2/3}} \left( \gamma_0 + \frac{1}{2} \big( (\chi_{HP} - \chi_{HS})(\psi_P - \varphi_P) + \chi_{HS}\varphi_H \big) \right), \tag{S28}$$

where $\rho_S$ is the number of macromolecules in the lamella per unit area. The free energy of the outer adjacent layers

$$\tilde{F}_{layer} = k_B T \frac{d_{out}}{\rho_S N v_S} (1 - \psi_P) \ln(1 - \psi_P), \tag{S29}$$

where the most favorable value of the distance $d_{out}$ is equal to $2l$, that permits to avoid the excluded volume interactions between the polar groups of different lamellas.

The volume fraction of hydrophobic groups

$$\varphi_H = \frac{\rho_S N}{d} v_H \tag{S30}$$

The volume fraction of polar groups inside and outside the lamellas

$$\varphi_P = f\varphi_H \frac{v_P}{v_H}\left(1 - \frac{2l}{d}\right), \quad \psi_P = f\varphi_H \frac{v_P}{v_H} \quad \text{for} \quad d > 2l, \tag{S31}$$

$$\varphi_P = 0, \quad \psi_P = f\varphi_H \frac{v_P}{v_H} \frac{d}{2l} \quad \text{for} \quad d < 2l.$$

The free energy of the lamellas with the middle layer of hydrophobic blocks contains additionally the contribution of the bulk interactions in the middle layer and the contribution of the interfaces between the middle layer and the layers of amphiphilic blocks, as for the core-shell micelles:

$$\tilde{F}_{Pl} = \tilde{F}_{bulk0} + \tilde{F}_{bulk} + \tilde{F}_{inter} + \tilde{F}_{surf} + \tilde{F}_{layer}, \quad (S32)$$

where the bulk free energies of the middle layers, $\tilde{F}_{bulk0}$, and the layers of amphiphilic blocks, $\tilde{F}_{bulk}$, are given by Eqs. (S4) and (S5), respectively. The contributions $\tilde{F}_{surf}$ and $\tilde{F}_{layer}$ are given by the equations (S28) and (S29). The interfacial contribution per monomer unit can be represented as

$$\tilde{F}_{inter} = k_B T \frac{\varphi_{H0} - \varphi_H}{\rho_S N v_S^{2/3}} \left( \gamma_0 + \frac{1}{2} \left( (\chi_{HP} - \chi_{HS}) \varphi_P + \chi_{HS} (\varphi_{H0} - \varphi_H) \right) \right). \quad (S33)$$

The volume fractions of hydrophobic groups:

$$\varphi_{H0} = \frac{\rho_S N_H}{d_0} v_H, \quad \varphi_H = \frac{\rho_S N_A}{d - d_0} v_H. \quad (S34)$$

The volume fractions of polar groups:

$$\varphi_P = \varphi_H \frac{v_P}{v_H} \left( 1 - \frac{2l}{d} \right), \quad \psi_P = \varphi_H \frac{v_P}{v_H} \quad \text{for} \quad d - d_0 > 2l, \quad (S35)$$

$$\varphi_P = 0, \quad \psi_P = \varphi_H \frac{v_P}{v_H} \frac{d - d_0}{2l} \quad \text{for} \quad d - d_0 < 2l.$$

From the incompressibility conditions, the solvent volume fractions $\varphi_{S0} = 1 - \varphi_{H0}$, $\varphi_S = 1 - \varphi_H - \varphi_P$.

By minimizing the free energies $\tilde{F}_{Pl0}$ (Eq. (S27)) and $\tilde{F}_{Pl}$ (Eq. (S32)) with respect to the surface density $\rho_S$ and the lamella thicknesses, the equilibrium parameters of the layered precipitates are calculated. It is worth noting that the free energy of the layered precipitate is asymptotically equal to that of discs with $r \to \infty$ or vesicles with $R \to \infty$.

For the precipitates with cylindrical and spherical cavities, we consider a simple case of the purely hydrophobic continuous domain surrounding the cavities of radius $R$ with the polar groups. The number of macromolecules per spherical cavity or per part of the cylindrical cavity of height $H$ is denoted by $m$. For the regular copolymer, this continuous domain is assumed to be homogeneous with the volume fraction of hydrophobic groups $\varphi_H$. For the block copolymer, the domain consists of a subdomain containing only hydrophobic blocks and layers of thickness $\Delta R$ of amphiphilic blocks surrounding the cavities. It is assumed that these layers are not overlapped. This condition is satisfied if the distance $d$ between the centres of the axes of the nearest cylinders or between the centres of the nearest cavities $d > 2(R + \Delta R)$ (Fig. S5). The volume fractions of hydrophobic groups in the subdomain and in the layers are $\varphi_{H0}$ and $\varphi_H$, respectively.

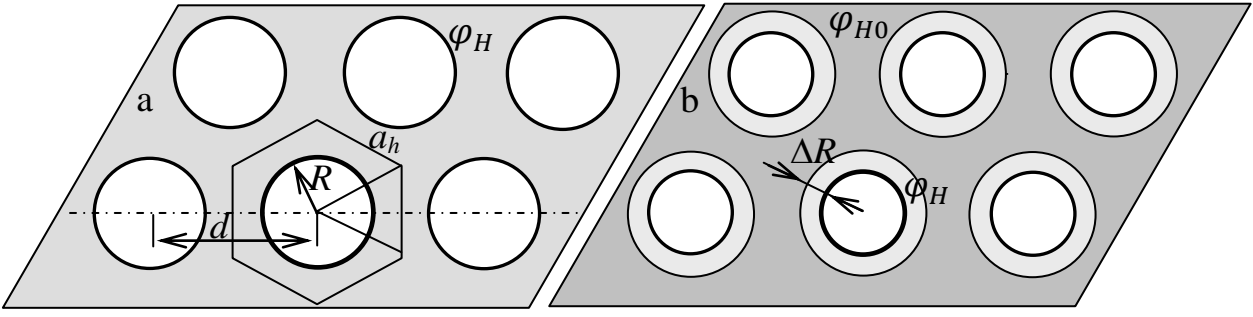

Figure S5. Scheme of the precipitates with cavities for (a) the homogeneously condensed polymer (Pc0 and Ps0) and (b) the condensed polymer forming the subdomain of hydrophobic blocks and the layers of amphiphilic blocks around the cavities (Pc and Ps). All polar groups are assumed to be in the cavities.

Since it is expected that a large number of cavities can be favourable, close-packed configurations are considered: hexagonally arranged cylinders and spheres with centers forming a hexagonal close-packed lattice. The scheme (Fig. S5) shows a cross-section by a plane perpendicular to the axes of the cylinders (or by the plane in which the centers of hexagonally arranged spheres lie). This plane can be divided into a system of hexagons with a side $a_h = d/\sqrt{3}$.

For the regular copolymer (Fig. S5a), the free energies per monomer unit for the morphologies with cylindrical or spherical cavities, $\tilde{F}_{Pc0}$ and $\tilde{F}_{Ps0}$ respectively, consist of the same contributions $\tilde{F}_{bulk}$, $\tilde{F}_{surf}$, and $\tilde{F}_{layer}$ as for the lamellas (Eq. (S27)). The bulk free energy $\tilde{F}_{bulk}$ is

$$\tilde{F}_{bulk} = k_B T \left( \chi_{HS}(1 - \varphi_H) + \frac{v_H}{v_S} \frac{1 - \varphi_H}{\varphi_H} \ln \varphi_S \right). \tag{S36}$$

The surface free energy is equal to

$$\tilde{F}_{surf} = k_B T \frac{\varphi_H}{\rho_S N v_S^{2/3}} \left( \gamma_0 + \frac{1}{2} \left( (\chi_{HP} - \chi_{HS}) \psi_P + \chi_{HS} \varphi_H \right) \right), \tag{S37}$$

where $\rho_S$ is the number of macromolecules per unit area of the cavity surface: $\rho_S = m/S_{cav}$, $S_{cav} = 2\pi R H$ for cylinders and $S_{cav} = 4\pi R^2$ for spheres. The free energy of the solution in the cavities (per monomer unit) can be represented as

$$\tilde{F}_{layer} = k_B T \frac{l_{cav}}{\rho_S N v_S} (1 - \psi_P) \ln(1 - \psi_P), \quad l_{cav} = \frac{V_{out}}{S_{cav}}, \tag{S38}$$

where $V_{out}$ is the volume of the region where the polar groups are localized, $\psi_P = fmN v_P/(MV_{out}) = f\rho_S N v_P/l_{cav}$ is the volume fraction of polar groups in this region. If $R < l$, then the polar groups are present throughout the entire cavity of volume $V_{cav}$: $V_{out} = V_{cav} = \pi R^2 H$ for cylindrical cavities and $4\pi R^3/3$ for spherical cavities ($l_{cav} = R/2$ and $R/3$, respectively). If $R > l$, then the polar groups are localized in the layer of thickness $l$ adjacent to the cavity surface: $V_{out} = \pi(R^2 - (R-l)^2)H$, $l_{cav} = l(1 - l/(2R))$ for cylindrical cavities and $V_{out} = 4\pi(R^3 - (R-l)^3)/3$, $l_{cav} = l(1 - l/R + l^2/(3R^2))$ for spherical cavities.

Let $\varphi_{cav}$ be the volume fraction of cavities in the precipitate: $\varphi_{cav} = V_{cav}/V_{sol}$, where $V_{sol}$ is the solution (or precipitate) volume per cavity. The value of $\varphi_{cav}$ is determined by the cavity radius $R$ and the distance $d$ (Fig. S5). For hexagonally arranged cylinders, the volume fraction $\varphi_{cav}$ is equal to the fraction of the circle area in a hexagon of area $S_h = 3\,da_h/2 = \sqrt{3}d^2/2$ (Fig. S5a):

$$\varphi_{cav} = \frac{\pi R^2}{S_h} = \frac{2\sqrt{3}\pi R^2}{3d^2}. \tag{S39}$$

For spheres in a hexagonal close-packed lattice, $\varphi_{cav}$ is the fraction of the sphere volume in a body bounded by the twelve planes perpendicular to the segments (at the midpoints) connecting a given vertex with the nearest ones. The volume of this body $V_{cpl} = \frac{4}{3}S_h H = \frac{\sqrt{2}}{2}d^3$, where $H = \frac{\sqrt{6}}{4}d$ is the distance between nearest planes with hexagonally arranged vertices. Then, the cavity volume fraction ($V_{sol} = V_{cpl}$)

$$\varphi_{cav} = \frac{4\pi R^3}{3V_{cpl}} = \frac{4\sqrt{2}\pi R^3}{3d^3}. \tag{S40}$$

The volume fraction $\varphi_H$ of hydrophobic groups is equal to $\varphi_H = mN\upsilon_H/(V_{sol}(1-\varphi_{cav}))$, it is determined by the cavity radius $R$, the distance $d$, and also the surface density $\rho_S$. For cylindrical cavities,

$$\varphi_H = \frac{4\sqrt{3}\pi R}{3(1-\varphi_{cav})d^2}\rho_S N \upsilon_H,$$

where the volume fraction of the cavity $\varphi_{cav}$ is given by Eq. (S39). For spherical cavities, the volume fraction of hydrophobic groups

$$\varphi_H = \frac{4\sqrt{2}\pi R^2}{(1-\varphi_{cav})d^3}\rho_S N \upsilon_H,$$

where the cavity volume fraction is given by Eq. (S40).

All polar groups are able to be in the cylindrical cavities if the hexagon side $a_h = d/\sqrt{3}$ is less than $R + l$, for the spherical cavities, the maximum distance between the body center and its vertex $a_{sph} = \sqrt{a_h^2 + \left(\frac{2H}{3}\right)^2} = d/\sqrt{2}$ should be less than $R + l$. The equilibrium values of the free energies of the precipitates with cavities formed by the regular copolymer and the geometrical characteristics $R$ and $d$ are calculated by minimizing $\tilde{F}_{pr(cyl0)}$ and $\tilde{F}_{pr(sph0)}$ with respect to the surface density $\rho_S$ and the sizes $R$ and $d$.

For the block copolymer, the hydrophobic groups of amphiphilic blocks are assumed to form layers of thickness $\Delta R$ adjacent to the cavities (Fig. S5b). All polar groups are able to be in the cavities if $\Delta R < l$. The free energies (per monomer unit) of the morphologies with cylindrical or spherical cavities, $\tilde{F}_{Pc}$ or $\tilde{F}_{Ps}$ respectively, consist of the same terms as for the lamellas (Eq. (S32)). The bulk free energies of the domains of hydrophobic and amphiphilic blocks are

$$\tilde{F}_{\text{bulk0}} = k_B T (1-f) \left( \chi_{HS}(1-\varphi_{H0}) + \frac{v_H}{v_S} \frac{1-\varphi_{H0}}{\varphi_{H0}} \ln \varphi_{S0} \right), \tag{S41}$$

$$\tilde{F}_{\text{bulk}} = k_B T f \left( \chi_{HS}(1-\varphi_H) + \frac{v_H}{v_S} \frac{1-\varphi_H}{\varphi_H} \ln \varphi_S \right). \tag{S42}$$

The contributions $\tilde{F}_{\text{surf}}$ and $\tilde{F}_{\text{layer}}$ are given by Eqs. (S37) and (S38), respectively. The interfacial contribution $\tilde{F}_{\text{inter}}$ is

$$\tilde{F}_{\text{inter}} = k_B T \frac{\varphi_{H0} - \varphi_H}{\tilde{\rho}_S N v_S^{2/3}} \left( \gamma_0 + \frac{1}{2} \chi_{HS} (\varphi_{H0} - \varphi_H) \right), \tag{S43}$$

where $\tilde{\rho}_S = m/\tilde{S}_{\text{int}}$, $\tilde{S}_{\text{int}}$ is the interface area of the layer of amphiphilic blocks of radius $R + \Delta R$: $\tilde{S}_{\text{int}} = 2\pi(R+\Delta R)H$, $\tilde{\rho}_S = \rho_S/(1+\Delta R/R)$ for cylinders and $\tilde{S}_{\text{int}} = 4\pi(R+\Delta R)^2$, $\tilde{\rho}_S = \rho_S/(1+\Delta R/R)^2$ for spheres.

The volume fractions of hydrophobic groups are equal to

$$\varphi_{H0} = \frac{(1-f)mNv_H}{V_{\text{sol}}(1-\varphi_{\text{cav}}) - V_{\Delta R}}, \quad \varphi_H = \frac{fmNv_H}{V_{\Delta R}}, \tag{S44}$$

where $V_{\Delta R}$ is the volume of the layer of thickness $\Delta R$ adjacent to the cavity. For cylinders, the cavity volume fraction is given by Eq. (S39) and the layer volume is $V_{\Delta R} = 2\pi R \Delta R_{\text{cyl}} H$, $\Delta R_{\text{cyl}} = \Delta R \left(1 + \frac{\Delta R}{2R}\right)$, then

$$\varphi_{H0} = \frac{(1-f)\rho_S N v_H}{(1-\varphi_{\text{cav}})\frac{\sqrt{3}d^2}{4\pi R} - \Delta R_{\text{cyl}}}, \quad \varphi_H = \frac{f \rho_S N v_H}{\Delta R_{\text{cyl}}}. \tag{S45}$$

For spheres, the cavity volume fraction is given by Eq. (S40) and the layer volume is $V_{\Delta R} = 4\pi R^2 \Delta R_{\text{sph}}$, $\Delta R_{\text{sph}} = \Delta R \left(1 + \frac{\Delta R}{R} + \frac{\Delta R^2}{3R^2}\right)$, then

$$\varphi_{H0} = \frac{(1-f)\rho_S N v_H}{(1-\varphi_{\text{cav}})\frac{\sqrt{2}d^3}{8\pi R^2} - \Delta R_{\text{sph}}}, \quad \varphi_H = \frac{f \rho_S N v_H}{\Delta R_{\text{sph}}}. \tag{S46}$$

Since we consider for simplicity only the case with all polar groups in the cavities, the free energies of a homogeneously condensed polymer (Fig. S5a) and a condensed polymer with a subdomain of hydrophobic blocks (Fig. S5b) are equal to each other if $\varphi_{H0} = \varphi_H$ under the same sizes $R$ and $d$. However, even in this case, the layered morphology can be realized for a wider range of possible $R$ and $d$ values than the homogeneously condensed polymer morphology. The equilibrium values of the free energy of the precipitates with cavities formed by the block copolymer and their geometrical characteristics are calculated by minimizing the expressions for $\tilde{F}_{\text{Pl}}$, $\tilde{F}_{\text{Pc}}$, and $\tilde{F}_{\text{Ps}}$ with respect to the surface density $\rho_S$ and the sizes $R$, $d$, and $\Delta R$.